\documentclass[twocolumn]{aastex62}

\newcommand\Gaia{\textit{Gaia}}

\newcommand\changed[1]{\textcolor{black}{#1}}

\usepackage{xcolor}
\usepackage{amsmath,amssymb,gensymb,times,graphicx,morefloats,color}
\usepackage{amsmath}
\usepackage{graphics,graphicx,url,verbatim,xspace}
\usepackage{verbatim}
\usepackage{changepage}
\usepackage{mathtools} 
\usepackage{comment}
\usepackage{gensymb}
\usepackage{afterpage}
\usepackage{hyperref}
\graphicspath{{./}{figures/}}

\received{January 4, 2020}
\revised{March 20, 2020}
\accepted{March 24, 2020}
\submitjournal{ApJ}

\shorttitle{Orbit fitting with \Gaia}
\shortauthors{Pearce et al.}

\begin{document}
\defcitealias{Dupuy2015OrbitalKepler-444A}{D16}

\title{Orbital Parameter Determination for Wide Stellar Binary Systems in the Age of \Gaia}

\correspondingauthor{Logan A. Pearce}
\email{loganpearce1@email.arizona.edu}

\author[0000-0003-3904-7378]{Logan A. Pearce}
\affil{Steward Observatory, University of Arizona, Tucson, AZ 85721, USA}
\affil{Department of Astronomy, University of Texas at Austin, Austin, TX, 78712, USA}
\affil{NSF Graduate Research Fellow}

\author[0000-0001-9811-568X]{Adam L. Kraus}
\affil{Department of Astronomy, University of Texas at Austin, Austin, TX, 78712, USA}

\author[0000-0001-9823-1445]{Trent J. Dupuy}
\affil{Gemini Observatory, Northern Operations Center, 670 N. A'ohoku Place, Hilo, HI 96720, USA}

\author[0000-0003-3654-1602]{Andrew W. Mann}
\affiliation{Department of Physics and Astronomy, The University of North Carolina at Chapel Hill, Chapel Hill, NC 27599, USA} 

\author[0000-0003-4150-841X]{Elisabeth R. Newton}
\affiliation{Department of Physics and Astronomy, Dartmouth College, Hanover, NH 03755, USA}

\author[0000-0003-2053-0749]{Benjamin M. Tofflemire}
\affil{Department of Astronomy, University of Texas at Austin, Austin, TX, 78712, USA}

\author[0000-0001-7246-5438]{Andrew Vanderburg}
\affil{Department of Astronomy, University of Texas at Austin, Austin, TX, 78712, USA}
\affiliation{NASA Sagan Fellow}

\begin{abstract}

The orbits of binary stars and planets, particularly eccentricities and inclinations, encode the angular momentum within these systems. Within stellar multiple systems, the magnitude and (mis)alignment of angular momentum vectors among stars, disks, and planets probes the complex dynamical processes guiding their formation and evolution. \changed{The accuracy of the \textit{Gaia} catalog can be exploited to enable comparison of binary orbits }with known planet or disk inclinations without costly long-term astrometric campaigns. 
We show that \textit{Gaia} astrometry can \changed{place meaningful limits on} orbital elements in cases with reliable astrometry, and discuss metrics for assessing the reliability of \textit{Gaia} DR2 solutions for orbit fitting.  We demonstrate our method by determining orbital elements for three systems (DS Tuc~AB, GK/GI~Tau, and Kepler-25/KOI-1803) using \textit{Gaia} astrometry alone.  We show that DS~Tuc~AB's orbit is nearly aligned with the orbit of DS~Tuc~Ab, GK/GI~Tau's orbit \changed{might be} misaligned with their respective protoplanetary disks, and the Kepler-25/KOI-1803 orbit is not aligned with either component's transiting planetary system.  We also demonstrate cases where \textit{Gaia} astrometry alone \changed{fails to provide useful constraints} on orbital elements. To enable broader application of this technique, \changed{we introduce the python tool \texttt{lofti\_gaiaDR2} to allow users to easily determine orbital element posteriors}.  

\end{abstract}


\section{Introduction} \label{sec:intro}

The monitoring of orbits is one of the oldest tools used to measure the properties and evolution of astrophysical systems. Orbital periods and semi-major axes convey the masses of systems, and have been used to weigh the universe in the context of planets (\citealt{Christy1978Pluto}), stars \citep{Aitken1918BinaryStars}, and galaxies and their dark matter halos (\citealt{RubinFord1970}). However, a broad range of astrophysical effects can also be probed using the other orbital elements-- eccentricities and orientations--that encode information about the magnitude and direction of the angular momentum vector(s) in a system. Even in its simplest form, much can be inferred about a system's formation and past history by the relative (mis)alignment of angular momentum vectors for different objects and on different scales.

On the scale of stellar systems (single or multiple) and their attendant planetary systems, angular momentum vectors trace their condensation out of interstellar clouds, and their subsequent orbital evolution through N-body interactions. In the classical picture of star formation, the collapse of a spherical protostellar core forms a central star whose rotation, circumstellar disk, and natal planetary system are aligned with the initial angular momentum of the primordial core (e.g., \citealt{Shu1987}). However, asymmetry in the mass distribution and gas motions of protostellar cores can be driven by phenomena like turbulence and magnetic fields, which complicate this simple picture of angular momentum evolution. \citep[e.g.,][]{BossBodenheimer1979AsymmetricCloudCollapse, Lee2019MultipleMagnetizeClouds}.

These same effects are thought to be the main drivers of wide ($a \ga 50$ AU; \citealt{Duchene2013Multiplicity, MoeDiStefano2017PsandQs}) binary star formation in the core fragmentation scenario \citep{BurkertBodenheimer1993MultipleFragCollapsingProtostars,Bate2000PropertiesofBinaries,Offner2010BinaryTurbulentFrag,Offner2016MisalignmentInMultiple}, which must operate broadly to account for the observed ubiquity of binary and higher order systems at these separations \citep[e.g.,][]{Duquennoy1991MultOfSolarTypeStars,Raghavan2010,Kraus2008MappingScorpius, Kraus2011MappingTaurus}. The subsequent evolution of protoplanetary disks in the binary environment can be perturbed from their initial configuration by torques from the companion \citep{Papaloizou1995,Batygin2012PrimordialMisalignments,Lai2014StarDiskBinaryInteractions}, while also dampening misalignment of orbiting bodies \citep{GoldreichTremain1980DiskSatelliteInteractions,Artymowicz1992BinaryPlanetaryDsikDynamics}.
These complex dynamical interactions can destroy or re-introduce alignment between planetary systems, binary orbits, and stellar rotation.  Additionally, binaries on scales with $a \la 100$ AU can also be formed via disk fragmentation, which would be expected to form well aligned systems \citep{Tobin2016diskfrag, TokovininMoe2019discfrag}.
Finally, discrete objects within the system can interact through long-term secular effects \citep[e.g.,][]{Kozai1962,Lidov1962,FabryckyTremain2007KozaiTidalFriction} or strong scattering via three-body interactions \citep{RasioFord1996DynInstabilities,WeidenschillinMarzari1996GravScattering,Bate2002CloseBinaryFormation,Chatterjee2008PlanetPlanetScatter} to exchange orbital energy and angular momentum, driving the evolution and randomization of angular momentum vectors.

There is extensive observational evidence for both alignment and misalignment among the stars, disks, and planets in binary systems. In the case of stellar properties, measurements for close systems ($a \la 50$ AU) suggest broad alignment of stellar spin \citep{Hale1994} and binary orbits \citep{TokovininLatham2017}, as well as for inner and outer orbits of hierarchical, high-order multiple systems \citep{Tokovinin2018DancingTwins}. In contrast, very wide systems do not show any correlation, though the ability to conduct this measurement is still quite new \citep[e.g.,][]{Tokovinin2018b}. Protoplanetary disks within young binary systems are seen in both states (mis-/aligned) with respect to each other (\citealt{Stapelfeldt1998HKTauBinaryDisk}; \citealt{Jensen2004ProtoplanetaryDiskAlignment}) and the binary orbit (\citealt{Winn2004PMSBinaryOccultation}; \citealt{Plavchan2013}; \citealt{Schaefer2014OrbitsPMS}). In systems which host circumbinary disks, binaries with $a \lesssim$ 1 AU exhibit tight alignment between the binary orbit and the disk plane, while systems with wider separations appear to have a random distribution of mutual inclinations \citep{Czekala2019}. The first orbital motion measurements for planet-hosting wide binaries rule out a random distribution of planet-binary mutual inclinations, showing that while the alignment might not be as strongly correlated as for multi-planet systems ($\la 5 \degr$; \citealt{Lissauer2011}), they tend towards alignment rather than random orientations with respect to the planets (e.g., \citealt{Dupuy2016Kepler444}; Dupuy et al., in prep).

The emerging trend from these studies is that the alignment between system components appears to increase with decreasing separation. This may be the effect of dynamical processes that align systems more efficiently at small separations, or a signpost for their formation process. In the case of planets, for instance, occurrence rates of protoplanetary disks and transiting planets are seen to decline for binary systems with separations less than $\sim$50 AU \citep{Krausetal2012,Krausetal2016}. Given the results above, this absence might suggest that mutual alignment is a condition for the systems (disk or planetary) that do survive. Testing this hypothesis, as well as determining the relevant scales where alignment becomes common/necessary (which appears to differ depending on the subsystem; binary-triple orbits, stellar rotation axes, binary-planetary orbits, etc) requires a large population of systems with well-characterized orbital parameters at a various evolutionary stages. While the sample is growing, constraining wide orbit systems is the bottleneck through which all future progress must pass.  

\changed{For decades there has been extensive effort to accurately determine probable orbits for visual stellar and substellar companion systems using time-series astrometric monitoring \citep[e.g.][]{Heintz1978DoubleStars,Kisselev1980AMPBinaryOrbits, Segransan2000AccurateMasses, Benedict2016MLRelation, Kiyaeva2017WideVisualBinaries}. These measurements have enabled comparisons with the known inclinations of planets or disks around individual components \citep[e.g.][]{Dupuy2016Kepler444,  Bryan2020ArxivObliquity}.}  The \Gaia\ astrometric revolution presents an opportunity to obtain \changed{similarly meaningful constraints on} wide stellar binary orbital parameters quickly in comparison.  \Gaia\ DR2 \citep{GaiaCollaboration2016TheMission, Gaia2018b} already reports precise positions and proper motions for many wide stellar binaries, and future data releases will increase the number of systems (resolving tighter pairs and presenting astrometry for fainter companions) and further improve the measurement uncertainties.  The wealth of new systems resolved by \Gaia\ is a boon to statistical studies of binary-disk and binary-planet alignment and permits new analyses of systems for which time series astrometry is not yet available.

\changed{As has long been noted for orbit fits, several families of orbit solutions match observations when a small fraction of an orbit is observed \citep{Aitken1918BinaryStars}.
In most cases, \Gaia\ measurements of relative positions and velocities do not provide enough observations to fully determine unique orbit solutions, however in this work we show that they can constrain the orbital elements to a scientifically useful degree.}
We find that for a range of binary configurations, \Gaia\ DR2 astrometry alone can provide orbital parameter posterior distributions that are consistent with, and in some cases superior to, long time series astrometric observations. We caution that not all \changed{binary systems} are amenable to this approach, as we discuss in Section 4.

In Section 2, we describe our orbit fitting technique combining Gaia DR2 relative positions and proper motions with the Orbits for the Impatient (OFTI, \citealt{Blunt2017OFTI}) rejection sampling orbit fitting method.  We also recommend metrics for assessing the quality of the \Gaia\ solution and accuracy of the technique.  In Section 3, we demonstrate the power of this technique by applying it to several binary systems resolved by \Gaia\ DR2.  In Section 4, we show how the quality of the \Gaia\ DR2 solution can limit the accuracy of this technique, and illustrate with examples where the technique fails to be as accurate as other methods.  In Section 5, we summarize the general classes of systems for which this technique is and is not sufficient, and discuss improvements with future \Gaia\ data releases.

\section{Method}
There are nine observable parameters describing the motion of one body relative to another --- three position ($X$, $Y$, $Z$), velocity ($\dot X$, $\dot Y$, $\dot Z$), and acceleration ($\ddot X$, $\ddot Y$, $\ddot Z$) terms, where $X$ and $Y$ define the plane of the sky.   Seven parameters are required to describe a Keplerian orbit in three dimensions --- semi-major axis ($a$), orbital period ($P$), eccentricity ($e$), inclination ($i$), argument of periastron ($\omega$), position angle of nodes ($\Omega$), and time of periastron passage ($T_0$).  \changed{If the total system mass is known, providing orbital period, then six independent measurements of the orbit are needed to obtain a unique orbit solution.  Traditional orbit determinations using time series astrometric measurements encode velocity and acceleration information in the time series data, and when a significant fraction of the orbit is observed, a unique orbit solution can be obtained.
\Gaia\ DR2 was obtained using time-series astrometry, but the catalog reports the binary relative $\Delta$RA ($\Delta\alpha$, corresponding to $Y$), $\Delta$DEC ($\Delta\delta$, $X$),  proper motion in RA/DEC ($\Delta\mu_{\alpha}$, $\dot Y$), $\Delta\mu_{\delta}$ $\dot X$), and in some cases radial velocity ($\Delta$RV, $\dot Z$) at a single epoch (2015.5 for Data Release 2; \citealt{Lindegren2018}). 
Thus four net observations, $X$, $Y$, $\dot X$, $\dot Y$, (or five, adding $\dot Z$ if radial velocities for both sources are present) can be used to constrain the orbit parameters at a single time point for a stellar binary for which both members are resolved by \Gaia. 
\Gaia\ parallaxes are not precise enough to constrain line of sight position ($Z$).}

\changed{The measurements reported in Gaia DR2 will not be sufficient to fully determine an orbit, and instead deliver a family of orbital solutions. In spite of this, we show in Section \ref{sec:validation} that the orbital constraints from Gaia can provide meaningful insight into system dynamics. The additional measurements reported in future data releases (such as acceleration terms) will further restrict those orbital solutions and increase the power of this technique; we describe the mathematical additions required to harness these acceleration measurements in Appendix A.3.}

\subsection{%The LOFTI 
Orbit fitting method}
\changed{The intent of this work is to examine the suitability of \Gaia\ astrometry as the sole measurement for constraining orbital elements of wide binaries. Any method of orbit fitting can be adapted to accommodate \Gaia\ position and velocity measurements.}
In this work, we have chosen to adapt the Orbits for the Impatient  \citep[OFTI, ][]{Blunt2017OFTI} rejection sampling algorithm to make use of the \Gaia\ astrometry.

Previously, OFTI has been used to compare astrometric observations, either separation/position angle (PA) or RA/DEC, at several observation epochs separated in time to predicted observations for trial orbits (\citealt{Blunt2017OFTI}, \citealt{Pearce2019GSC6214}).  Here we adapt OFTI to fit the linear plane-of-sky velocity vector provided by \Gaia\ DR2, and adopted the name LOFTI (Linear OFTI) for this application.  

Rather than time-series observations, our modified OFTI uses relative $\Delta\alpha$, $\Delta\delta$, relative proper motions ($\Delta\mu_{\alpha}$, $\Delta\mu_{\delta}$), and $\Delta$RV if available, at the single \Gaia\ DR2 epoch (2015.5) to constrain orbital parameters. Here $\Delta\mu_{\alpha}$ is chosen to be $\Delta\mu_{\alpha}$~=~$\Delta\mu_{\alpha, sec}~-~\Delta\mu_{\alpha, pri}$, likewise $\Delta\alpha$, $\Delta\delta$, $\Delta\mu_{\delta}$ and $\Delta$RV.

The OFTI method is described in detail in \cite{Blunt2017OFTI}, but we summarize briefly here.  OFTI is a rejection sampling algorithm that randomly generates four orbital parameters from uniform priors for eccentricity (\textit{e}), argument of periastron ($\omega$), mean anomaly from which we derive epoch of periastron passage (\textit{t$_0$}), and cosine of inclination cos(\textit{i})\footnote{In the absence of radial velocity information, a degeneracy exists between $\Omega$ and $\omega$.  In all orbital parameter posteriors reported in this work we restrict $\Omega$ to $[0,\pi]$ with the exception of the LOFTI fit to DS Tuc, which included radial velocity measurements for both objects, breaking the degeneracy.}.  \changed{We assume the total system mass from prior measurements in the literature (typically based on previous temperature and luminosity measurements) and utilize the \Gaia\ parallax in order to remove orbital period as a free parameter via Kepler's 3rd law.}  %The semi-major axis (\textit{a}) and longitude angle of ascending node ($\Omega$) for all trial orbits are initially fixed. 
\changed{The semi-major axis (\textit{a}) and longitude angle of ascending node ($\Omega$) for all trial orbits are scaled and rotated from an arbitrary initial value to match the \Gaia\ $\Delta\alpha$ and $\Delta\delta$ positions}.  The scale-and-rotate step speeds up the rejection sampling process by avoiding the large majority of potential orbits with extremely discrepant separations or PAs, \changed{and means that for our six-dimensional parameter space, we have only four free parameters}.  After scale-and-rotate, we compute the relative velocities of each trial orbit at the observation date, and perform a rejection sampling accept/reject decision, in which an orbit is accepted if its probability is larger than a random\changed{ly chosen} number from the interval [0,1].  The probability of a trial orbit is given as P$(trial \;orbit | observations) = e^{-\frac{\chi^2}{2}}$ where 

\begin{centering}
 \begin{align}
\chi^2 = %\big(\frac{\Delta\alpha - Y}{\sigma_{{\alpha}}}\big)^2 + \big(\frac{\Delta\delta - X}{\sigma_{{\delta}}}\big)^2 + \nonumber \\
\big(\frac{\Delta\mu_{\alpha} - \dot Y}{\sigma_{\mu_{\alpha}}}\big)^2 +  \big(\frac{\Delta\mu_{\delta} - \dot X}{\sigma_{\mu_{\delta}}}\big)^2
 \end{align}
\end{centering}
$X$, $Y$, $\dot X$, and $\dot Y$ denote predictions from trial orbits, while $\alpha,\, \delta,\, \mu_\alpha,\, \mu_\delta$ denote observations from \Gaia.  An additional radial velocity term can be added to the $\chi^2$ if applicable.

OFTI is particularly well-suited for poorly constrained orbital motion, such as long-period systems with astrometry for only a small orbit fraction, and is able to quickly determine orbital element posteriors for systems where Markov Chain Monte Carlo (MCMC) might not converge (\citealt{Blunt2017OFTI}, \citealt{Blunt2019orbitizeArxiv}).  However, it can be prohibitively slow when the orbit is well-constrained \changed{because the majority of trials will be rejected due to low probability} (\citealt{Blunt2019orbitizeArxiv}). Future \Gaia\ data releases will include more systems with well-determined radial velocities ($\dot Z$), and accelerations in the plane of the sky ($\ddot X$, $\ddot Y$), further constraining the orbit.  If more constraints are added, the orbit fit would be better served by another fitting method such as MCMC.  However, even in this case OFTI is still useful in narrowing the parameter space and determining a starting point for an MCMC.

\changed{The accuracy of OFTI depends on the accuracy of the astrometric observations and their assessed uncertainties, which we discuss in the following section. Furthermore, while OFTI fully samples the parameter space, the posteriors will still be subject to degeneracies among orbital parameters. Some pathological orbits, or orbits observed at pathological times, might be subject to irreducible degeneracies even in the case where the data are accurate and precise. For example, when the motion is purely along the separation direction, it can be difficult without additional information to distinguish between an edge-on orbit and a highly eccentric orbit. Finally, the need to estimate component stellar masses will introduce a dependence on stellar evolutionary models and external data sources, with all of the associated uncertainties associated with both. There may also be cases where systematic errors can emerge, such as when unidentified additional stellar components within the system contribute additional mass that is not incorporated into the model. These risks must be assessed on a case-by-case basis, but if they can be reduced to an acceptable level, then LOFTI provides a powerful tool for many astrophysical applications.
}

To allow users to easily and quickly implement this technique, we provide the simple python package \texttt{lofti\_gaiaDR2}\footnote{\url{https://github.com/logan-pearce/lofti\_gaiaDR2}}.  The GitHub repository includes documentation and examples for implementing this code, and Appendix 2 briefly describes the python tool.

\subsection{Data Quality Indicators}\label{sec:quality}

\changed{For any astrometric orbit-fitting technique, the utility and accuracy of the orbital solution will depend on the accuracy and precision of the observations.
For \Gaia\ measurements,} there are many potential quality indicators for a given solution \citep{Lindegren2018,Lindegren2018RUWE}, but the most complete indicator is the re-normalized unit weight error (RUWE) \citep{Lindegren2018RUWE}.   
In brief, RUWE is the square root of the reduced $\chi^2$ statistic, corrected for its dependence on color and magnitude. It can be computed using methods described in \cite{Lindegren2018RUWE}, or found in the table \texttt{gaiadr2.ruwe} of the \Gaia\ archive\footnote{\url{https://gea.esac.esa.int/archive/}}.  RUWE close to 1.0 indicates that the single star model is a good fit to observations; higher RUWE, such as RUWE $>$ 1.4, are typically found to be spatially resolved binaries with compromised astrometric accuracy \citep[e.g.][]{Rizzuto2018ZEIT,Ziegler2019SOARGaia_arxiv}.

Other useful metrics are available in the \Gaia\ DR2 catalog and discussed in the \Gaia\ DR2 documentation\footnote{\url{http://gea.esac.esa.int/archive/documentation/GDR2/}}.
%%%%%%%%%%%%%%%%%%%%%%%%%%%%

\begin{deluxetable*}{cccccccc}[htb!]
\tablecaption{{Selected \Gaia\ binaries and their input parameters}\label{table:Gaia solutions}}
%\tablewidth{0.45\textwidth}
\tablehead{ 
\colhead{Name} & \colhead{Mass} & \colhead{G mag} & \colhead{Parallax}  & \colhead{RUWE}   & \colhead{Separation [mas]} & \colhead{Velocity$^{j}$} &\colhead{WDS ID}\\
\colhead{\Gaia\ DR2 Source ID}  & \colhead{[M$_\odot$]} & \colhead{[mag]} & \colhead{[mas]}  &  \colhead{} & \colhead{P.A. [deg]} & \colhead{[km s$^{-1}$]} & \colhead{(if applicable)} 
%\colhead{WDS desig (if used) } & \colhead{}  & \colhead{} & \colhead{} & \colhead{}  &\colhead{} &\colhead{}
} 
\startdata
\textbf{DS Tuc A} & 1.01$\pm$0.06$^{a}$ & 8.32 & 22.666 $\pm$ 0.035 &  1.034 & 5364.61$\pm$0.03 & 1.94$\pm$0.72$^{k}$ & 23397-6912 \\
6387058411482257536 & &  &  &  & &  & \\
%\cline{1-5}
\rule{0pt}{3ex}  \textbf{DS Tuc B} & 0.84$\pm$0.06$^{a}$ & 9.40 & 22.650 $\pm$ 0.029 & 1.015 & 347.6582$\pm$0.0002\\
6387058411482257280 &  & &  &  &   \\
\hline
\textbf{GK Tau} &  0.79$\pm$0.07$^{b}$ &  11.88 & 7.7362 $\pm$ 0.0434  & 1.089 & 13157.97$\pm$0.05 & 1.15$\pm$0.11 & ... \\
147790206908395776 &  &  &  & &    \\
\rule{0pt}{3ex} \textbf{GI Tau}  & 0.53$^{+0.09\,c}_{-0.11}$ &  12.70 & 7.6625 $\pm$ 0.0460 &  1.046 & 328.4399$\pm$0.0002\\
147790202612482560 &   &  &  &   \\
\hline
\textbf{Kepler-25} & 1.159$^{+0.040\,d}_{-0.051}$& 10.63 & 4.082 $\pm$ 0.024  & 0.998 & 8415.29$\pm$0.02 & 0.879$\pm$0.081 & ... \\
2100451630105041152 &  & & &  \\
\rule{0pt}{3ex} \textbf{KOI-1803} & 0.794$\pm$0.026$^{e}$& 13.24 & 4.089 $\pm$ 0.014 &  1.060 & 288.2891$\pm$0.0002\\
2100451630105040256 &  & & &  \\
\hline
\textbf{Gl 896 A} & 0.3880$\pm$0.0091$^{f}$ & 9.04 & 159.710 $\pm$ 0.082 &  1.173  & 5380.71$\pm$0.09 & 2.477$\pm$0.009 & 23317+1956\\
2824770686019003904 &  &  &  &   \\
\rule{0pt}{3ex} \textbf{Gl 896 B} & 0.2519$\pm$0.0076$^{f}$ & 10.82 & 160.060 $\pm$ 0.108  & 1.460  & 78.0372$\pm$0.0008\\
2824770686019004032 &   &  &  &   \\
\hline
\textbf{Kepler-444 A} & 0.76$\pm$0.04$^{g}$ & 8.64 & 27.414 $\pm$ 0.029  & 1.000 & 1837.44$\pm$0.34 & 2.39$\pm$0.24 & ... %19190+4138AB
\\
 2101486923385239808 &  &  &  &   \\
\rule{0pt}{3ex} \textbf{Kepler-444 BC} &  0.54$\pm$0.04$^{h}$ & 12.27 & 32.652 $\pm$ 0.569  & 16.687 & 252.76$\pm$0.02 \\
2101486923382009472 &  &  &  &   \\
\hline
\textbf{RW Aurigae A}  & 1.20$\pm$0.24$^{i}$ & 11.11 & 6.1157 $\pm$ 0.067  & 1.774 & 1488.63$\pm$0.66 & 2.25$\pm$1.18 & ... %05078+3024A,BC
\\
156430822114424576 &  &  &  &   \\
\rule{0pt}{3ex} \textbf{RW Aurigae B}  & 0.72$\pm$0.14$^{i}$ & 11.40 & 6.583 $\pm$ 0.902  & 28.904 & 74.47$\pm$0.03 \\
156430817820015232 &  &  &  &   \\
\enddata
\tablecomments{All input parameters derived from \Gaia\ astrometry, and given as the second component relative to the first component.  (a) \citealt{Newton2019DSTuc}, (b) \citealt{Kenyon1995}, (c) \citealt{Akeson2019YoungBinaryDisks}, (d) \citealt{Aguirre2015Asteroseismology}, (e) \citealt{FultonCalifKeplerSurvey2018}, (f) this work, via the mass-luminosity relation of \citealt{Mann2019MassLuminosityRelation} and 2MASS K-band magnitudes, (g) \citealt{Campante2015}, (h) \citealt{Dupuy2016Kepler444}, (i) \citealt{Kraus2011MappingTaurus}, (j) Velocity in the plane of the sky, given as $v = \sqrt{ \mu_{\alpha}^2 + \mu_\delta^2}$   \;(k) DS Tuc relative velocity includes radial velocity.}
\end{deluxetable*}

\section{Validation demonstrated by successful fits}\label{sec:validation}

To assess the validity of the fitting method, we tested if a fit constrained by \Gaia\ astrometry would return the same posterior distribution as a multi-epoch astrometric OFTI fit for a given system (Section 3.1), and applied this technique to new interesting systems (Section 3.2 and 3.3).  Table \ref{table:Gaia solutions} lists the binary systems studied in this and Section \ref{sec:limitations}, and their input parameters derived from \Gaia\ astrometry.

\begin{deluxetable*}{cccccccccccc}[htb!]
\tablecaption{{Summary of Orbital Parameters for DS Tuc AB}\label{table:dstuc orbits}}
%\tablewidth{0.45\textwidth}
\tablehead{\colhead{} & \multicolumn{5}{c}{Astrometry-Only} & \colhead{} & \multicolumn{5}{c}{\Gaia\ Position/Velocity-Only} \\
\cline{2-6}
\cline{8-12}
\colhead{Element} & \colhead{Median} & \colhead{Std Dev} &\colhead{Mode} & \colhead{68.3\% Min CI} & \colhead{95.4\% Min CI} & \colhead{} & \colhead{Median} & \colhead{Std Dev} & \colhead{Mode} & 
\colhead{68.3\% Min CI} & \colhead{95.4\% Min CI}}
\startdata
\changed{log($a$)} (AU) & 2.35 & 0.30 & 2.38 & (2.08, 2.47) & (2.08, 3.12) & & 2.28 & 0.21 & 2.21 & (2.19, 2.36) & (2.19, 2.80) \\
%$P$ (yrs) & 8820 & 40060 & 1130 & (970, 3740) & (950, 36250) & & 1760 & 510 & 1500 & (1470,1730) & (1470,2440) \\
$e$ & 0.63 & 0.29 & 0.99 & (0.50, 0.99) & (0.09, 0.99) & & 0.57 & 0.10 & 0.47 & (0.46, 0.60) &   (0.46, 0.77)\\
$i$ (\degree) & 98.2 & 11.4 & 93.3 & (90.1, 98.1) & (86.8, 126.2) & & 96.9 & 0.9 & 96.6 & (96.0, 97.8) & (95.0, 98.6) \\
$\omega$ (\degree) & 6 & 103 & 302 & (-71, 156) & (-160, 180) & & 6 & 35 & 36 & (340, 52) & (297, 62)\\
$\Omega$ (\degree) & 66 & 95 & 348 & (-15, 168) & (-36, 179) & & 167 & 16 & 167 & (164, 170) & (163, 174)\\
$T_0$ (yr) & -2380 & 32970 & 1400 & (140, 1640) & (-13160, 2015) & & 1250 & 480 & 1520 & (1250, 1530) &  (-590, 1530)\\
%\hline
\changed{log(Periastron)} (AU) & 1.90 & 0.79 & 2.35 & (1.13, 2.41) & (-0.09, 3.14) & & 1.91 & 0.21 & 1.94 & (1.73, 2.08) & (1.35, 2.17)\\
\enddata
%\tablecomments{}
\end{deluxetable*}

\begin{figure*}
\centering
\includegraphics[width=0.47\textwidth]{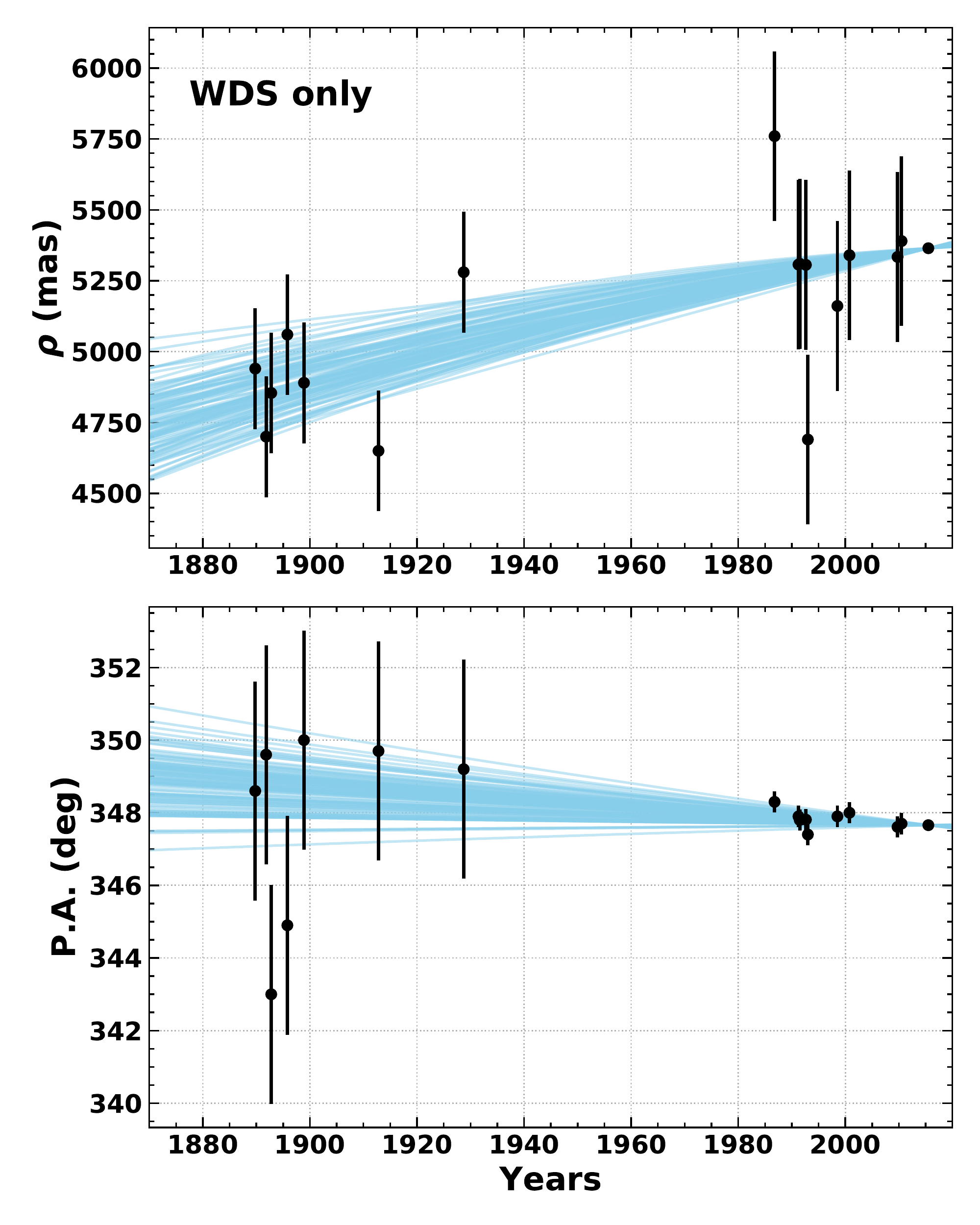}
\includegraphics[width=0.47\textwidth]{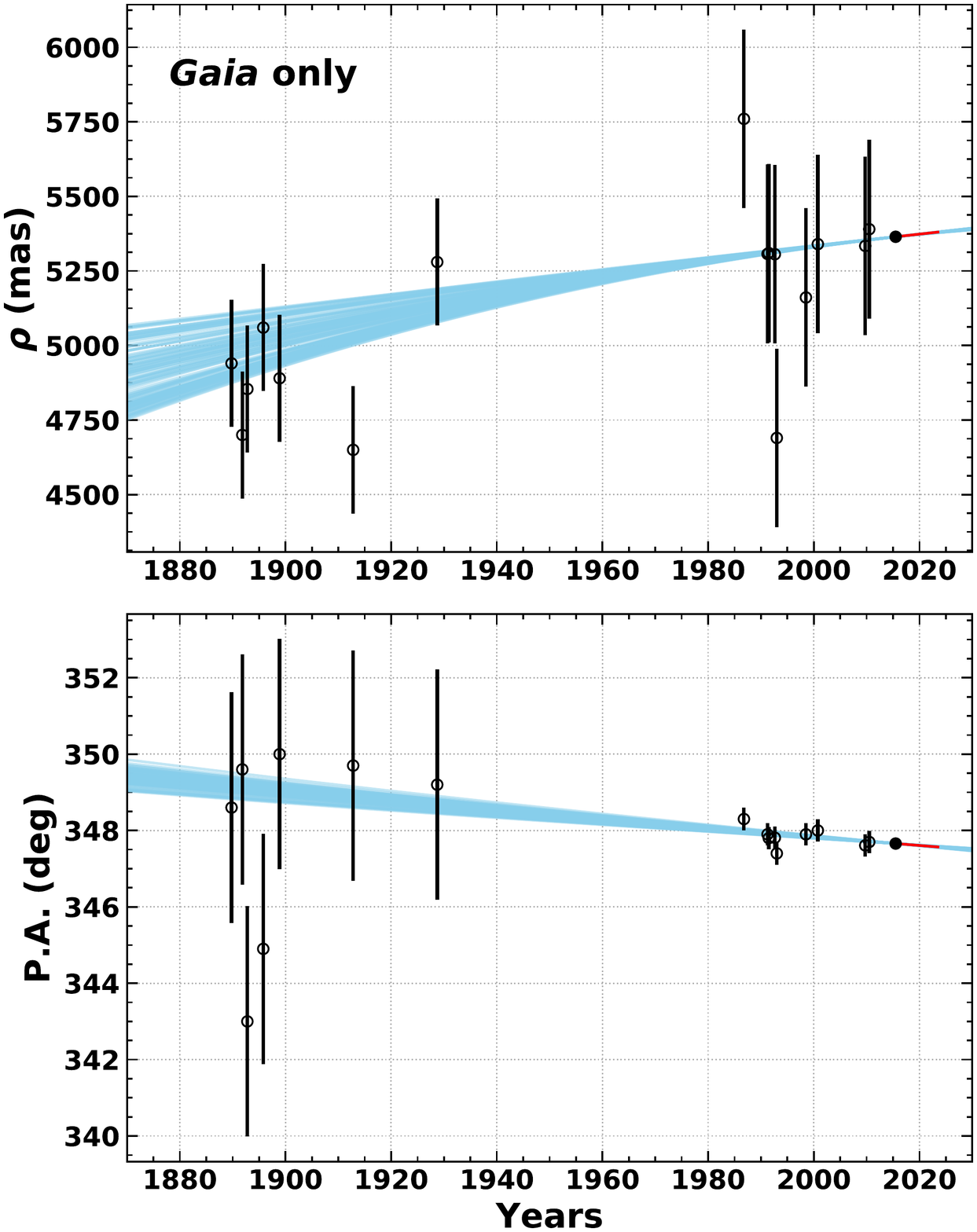}
\caption{\small{\textbf{Successful fit: DS Tuc}. Left: A selection of 100 orbits for DS~Tuc~B relative to DS~Tuc~A from the posterior of orbits, fit using the WDS astrometry points only.  The WDS astrometry data are overplotted.  Error bars are estimated from root mean square error about two linear fits to separation and position angle independently for the pre-1940 group and the post-1980 group.  Scatter, and thus error bars, are not the same in both dimensions, and so systematic error differences exist that are not accounted for in our error estimate.
Right: A selection of 100 orbits from the posterior of the fit against \Gaia\ DR2 position, velocities, and radial velocity, with the WDS astrometry overplotted for comparison (open circles, not used in fit).  The red line indicates the direction of the \Gaia\ velocity vector.  The \Gaia\-only fit matches the trend of the WDS-only fit, but is more tightly constrained.}}
\label{fig:dstuc astr fit}
\end{figure*}

\begin{figure*}
\centering
\includegraphics[width=0.49\textwidth]{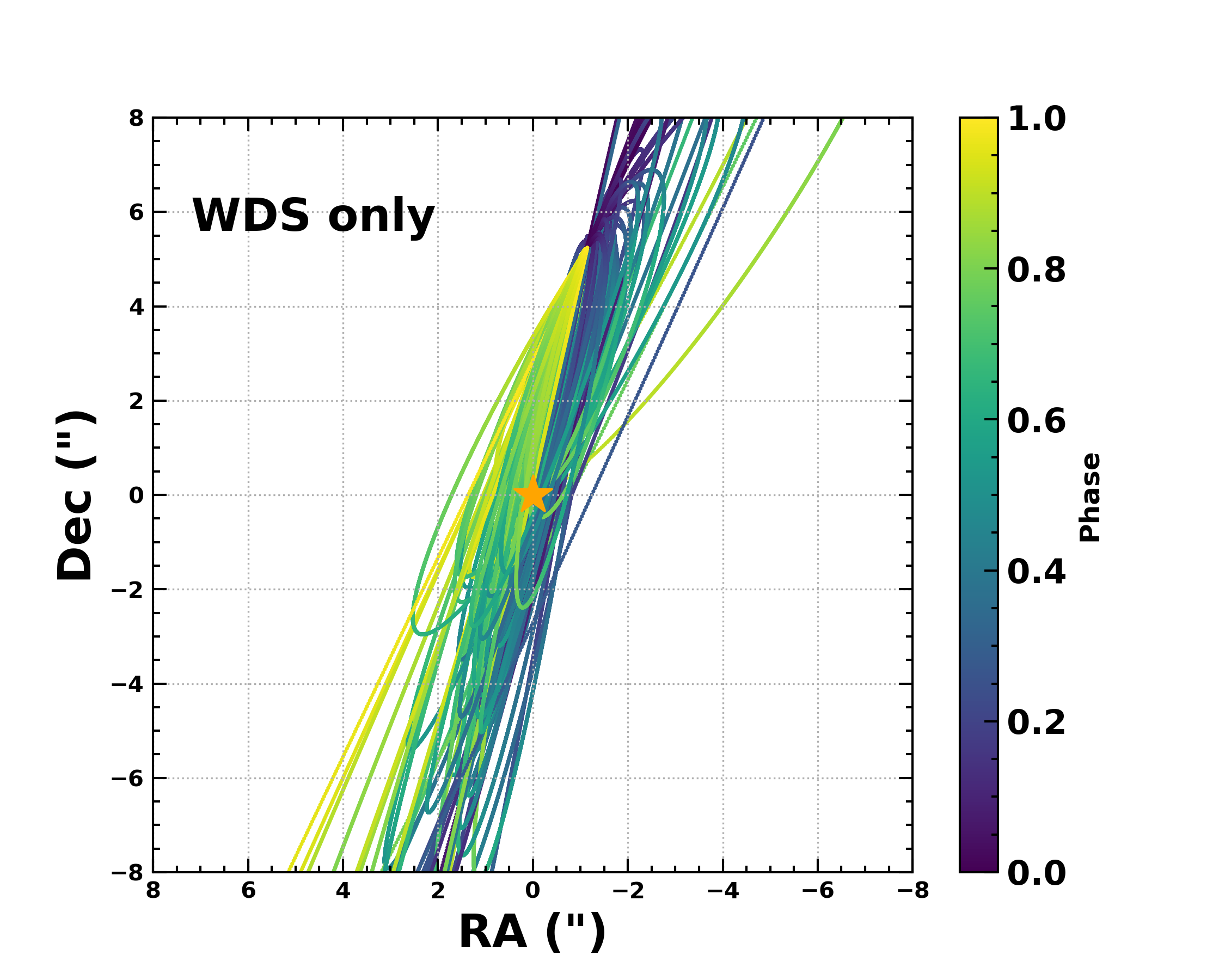}
\includegraphics[width=0.49\textwidth]{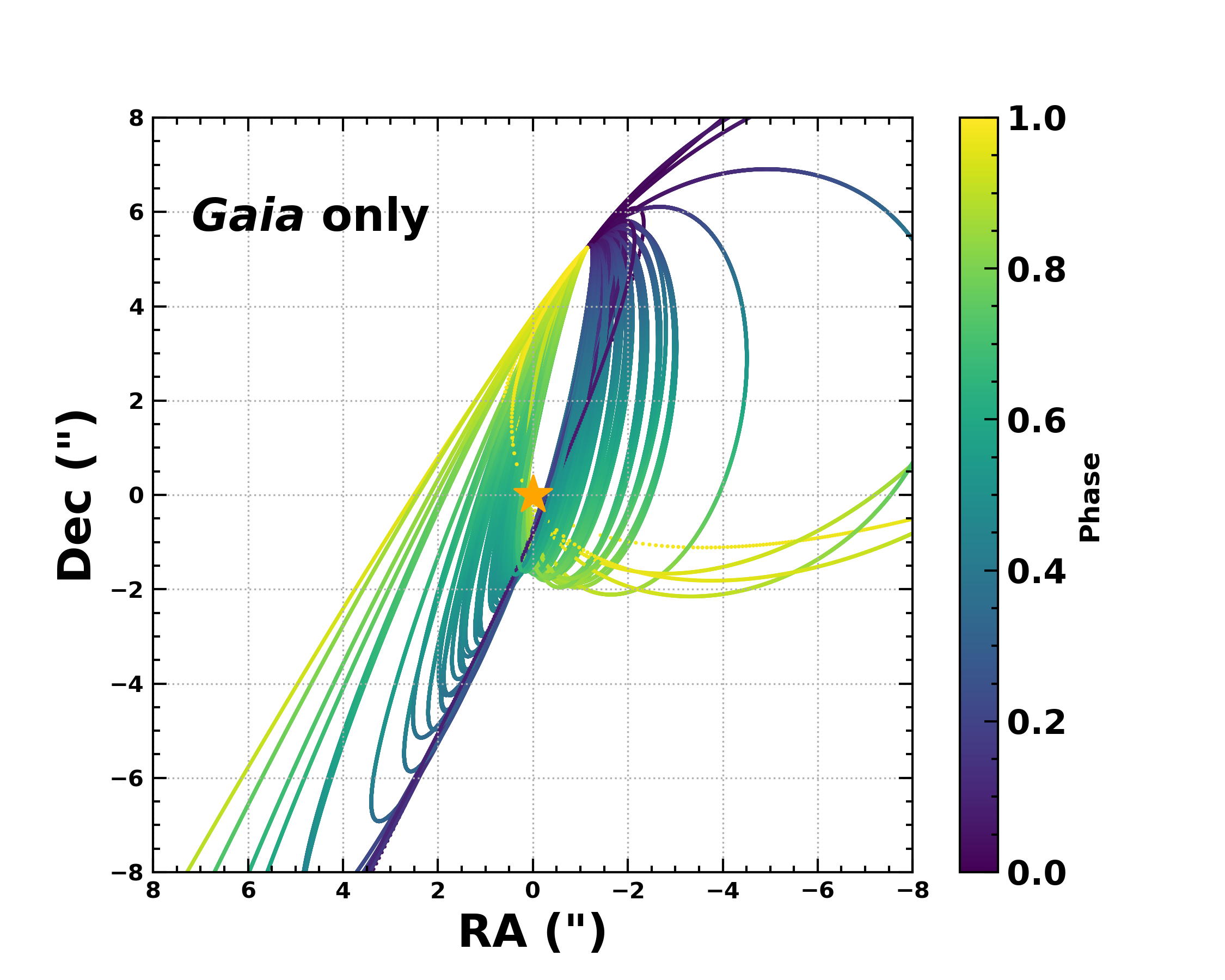}
\caption{\small{\textbf{DS Tuc}.  Left: Selection of 100 orbits for DS~Tuc~B relative to DS~Tuc~A from the posterior of orbits fit against the WDS astrometry data only  in the plane of the sky.  Right: Selection of 100 orbits from the posterior of the fit against \Gaia\ DR2 position, velocities, and radial velocity.}}
\label{fig:dstuc sky orbits}
\end{figure*}

\begin{figure*}
\centering
\includegraphics[width=0.95\textwidth]{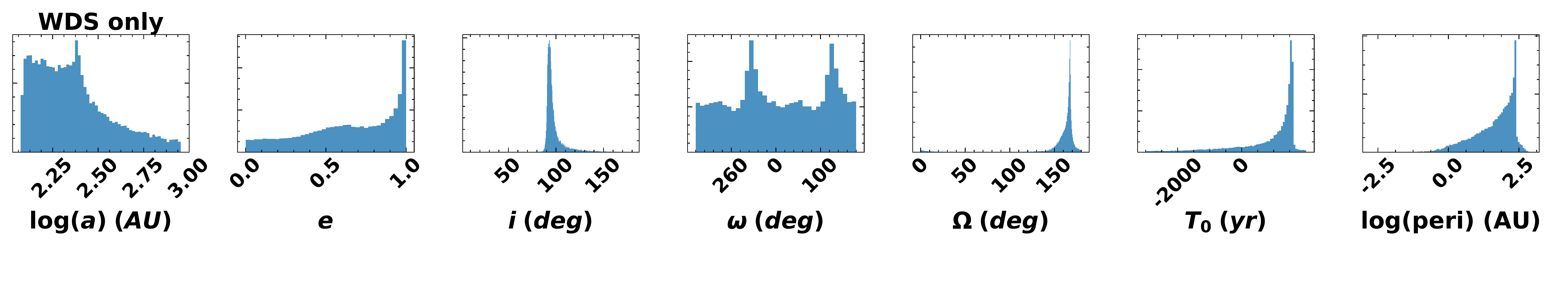}
\includegraphics[width=0.95\textwidth]{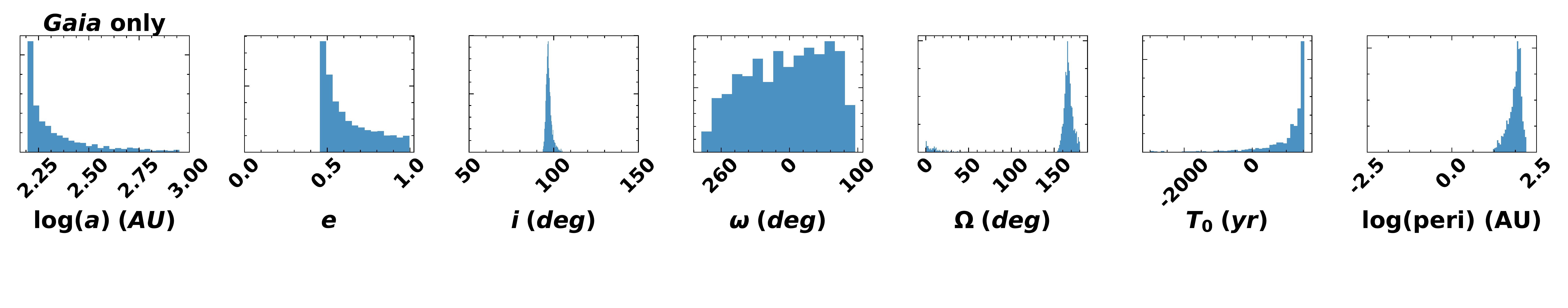}
\caption{\small{\textbf{DS Tuc}.  Top: Histogram of posterior distributions of orbital elements for the WDS astrometry-only OFTI fit.  Bottom: Histograms of orbital elements for the \Gaia\ linear velocity fit, including \Gaia\ radial velocities.  Semi-major axis (a) and epoch of periastron passage (T$_0$) have been truncated in both for clarity.  The linear fit agrees broadly with the astrometry-only fit, but is much more tightly constrained, and suppresses some of the more extreme orbital solutions, most notably circular and low eccentricity, very high eccentricity, and large semi-major axis orbits.}}
\label{fig:dstuc fit hists}
\end{figure*}

\begin{figure}
\centering
\includegraphics[width=0.5\textwidth]{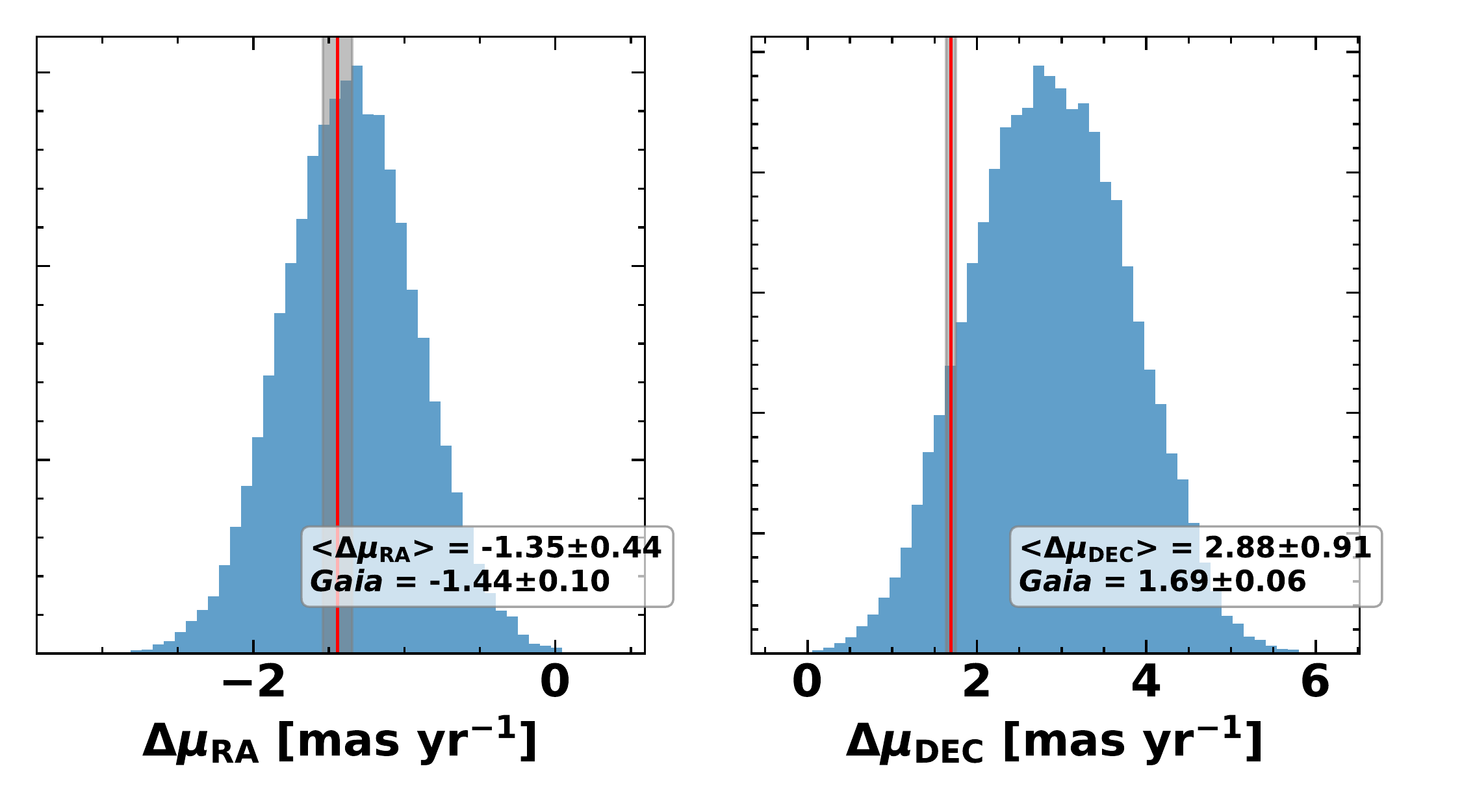}
\caption{\small{\textbf{DS Tuc}. Posterior distribution of instantaneous orbital velocity in right ascension (left; \changed{$\Delta\mu_{\rm{RA}}$}) and declination (right; \changed{$\Delta\mu_{\rm{DEC}}$}) from the long period astrometric (WDS only) DS Tuc orbit fit, with the \Gaia\ DR2 measurements indicated by a red vertical line and grey shading for the uncertainties.  The \Gaia\ measurement agrees with the mean of the astrometric posterior in right ascension, and is 1.3-$\sigma$ \changed{from the mean of the posterior distribution} in declination. }}
\label{fig:proper motion comparison}
\end{figure}

\subsection{DS Tuc with WDS and \Gaia}

The young planet host DS Tuc (HD 222259) is a visual binary \citep{Torres1988VisualBinariesIII} with DS Tuc A (Spt = G6V, M$_*$ =  1.01 $\pm$ 0.06 M$_{\odot}$; \citealt{Newton2019DSTuc}) and DS Tuc B (SpT = K3V, M$_*$ =  0.84 $\pm$ 0.06 M$_{\odot}$; \citealt{Newton2019DSTuc}) separated by 5$''$ \citep{Torres2006}.  The masses were determined using isochrones by Newton et al. DS Tuc A hosts a transiting planet with a radius of 5.70$\pm$0.17 R$_\earth$ \citep{Newton2019DSTuc}.  In \cite{Newton2019DSTuc}, we used LOFTI to study the orbital alignment of DS Tuc AB, and found the binary was nearly aligned with the planet orbit.  Here we demonstrate how our technique provided \changed{
%a smaller family of orbital solutions than did
a tighter constraint on most orbital elements than} time series astrometry alone for this system.

The Washington Double Star \citep[WDS, ][]{Mason2001WDS} Catalog provides separation and position angle measurements for stellar binaries spanning decades.  The WDS catalog provides a long time baseline for computing orbits astrometrically, which we verified against orbital posteriors for \Gaia\ DR2 linear velocity orbit fits.  WDS data for DS Tuc go back as far as 1870, and both objects have well-defined solutions in \Gaia\ DR2, including radial velocities.  We performed an astrometry-only fit to the WDS measurements with standard OFTI, and compared the results to a \Gaia\ LOFTI fit.

\textbf{Astrometric fit with standard OFTI:}  We fit orbital parameters to the relative astrometry of DS Tuc AB using the established OFTI method described in \cite{Pearce2019GSC6214} and \cite{Blunt2017OFTI}.  The WDS astrometry used in the fit is displayed in Figure \ref{fig:dstuc astr fit}.  To establish errors on separation and position angle we used the root-mean-squared error (RMSE) about a linear fit to the WDS astrometry.  As can be seen in Figure \ref{fig:dstuc astr fit}, the observations divide into two groups, an earlier group (before 1940) with more scatter, and a later group (after 1980) with smaller scatter, and so we determined the RMSE about that line for each group independently.  The latest data point is the \Gaia\ DR2 measurement, which uses the \Gaia\ reported error.    

\textbf{\Gaia\ position/velocity vector fit:} We performed a LOFTI position/velocity vector fit using the relative $\Delta \alpha$, $\Delta \delta$, $\mu_\alpha$, $\mu_\delta$, and radial velocity reported in \Gaia\ DR2.  The \Gaia\ relative radial velocity measurement $\Delta$RV = 1.88$\pm$0.72 km s$^{-1}$ agrees with the mean relative radial velocity measurement of $\Delta$RV = 1.64$\pm$0.35 km s$^{-1}$ reported by \cite{Newton2019DSTuc}.

The posterior parameters for both fits are given in Table \ref{table:dstuc orbits}.  Figures \ref{fig:dstuc astr fit} and \ref{fig:dstuc sky orbits} show a selection of orbits from the astrometric (WDS only, left) and velocity vector (\Gaia\ only, right) fits. Marginal posterior distributions for orbital parameters for each fit are shown in Figure \ref{fig:dstuc fit hists}.  

Figure \ref{fig:dstuc astr fit} demonstrates that both fits agree broadly with the astrometric points and display the same general trend, however the linear velocity fit is more tightly constrained.  The 1-dimensional orbital element distributions also broadly agree, but the linear velocity fit has suppressed some of the more extreme orbits, namely highly eccentric, larger semi-major axis, and extremely close periastron orbits.  Circular and low eccentricity orbits are also ruled out from the \Gaia\ fit. 

Finally, Figure \ref{fig:proper motion comparison} displays the posterior distribution of \changed{proper motion in} RA (\changed{$\mu_{\rm{RA}}$}) and Dec (\changed{$\mu_{\rm{DEC}}$}) \changed{computed from} posterior orbits in the WDS-only fit.  The \Gaia\ DR2 proper motions are indicated with a red vertical line.  The \Gaia\ measurements are in the posterior of the astrometry-only fit.  The mean of the  \changed{$\mu_{\rm{RA}}$} distribution agrees with \Gaia\ inside of 1$\sigma$, and the \changed{$\mu_{\rm{DEC}}$} mean is a little more than 1$\sigma$ from the \Gaia\ measurement.  This is likely the source of the differences in parameter posterior distributions between the astrometry and the \Gaia\-only fits, with the \Gaia\ measurement being more robust due to small fractional uncertainties.

We conclude from this analysis that for DS Tuc, fitting the \Gaia\ astrometry alone produced a posterior that agrees with, but is more precise than, fitting against time series astrometry alone.  Combining the time series astrometry with the \Gaia\ velocities resulted in an even more tightly constrained orbital parameter posterior, which is reported in \cite{Newton2019DSTuc} and is used in their analysis of this system.  %However, due to the tight constraints, OFTI is not efficient and the orbit fit is better served by an MCMC-based algorithm.

\begin{figure*}
\centering
\includegraphics[width=0.5\textwidth]{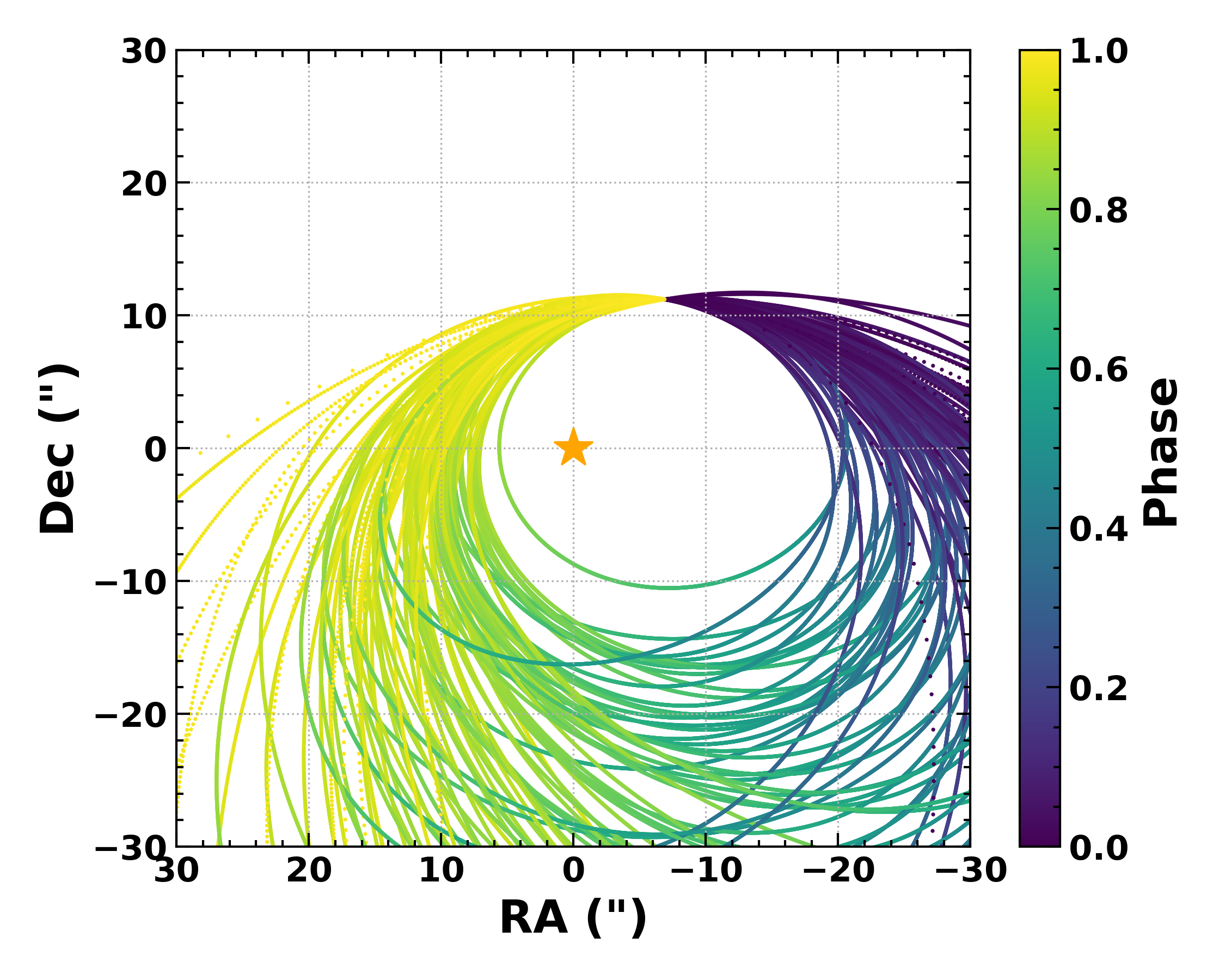}
\includegraphics[width=0.95\textwidth]{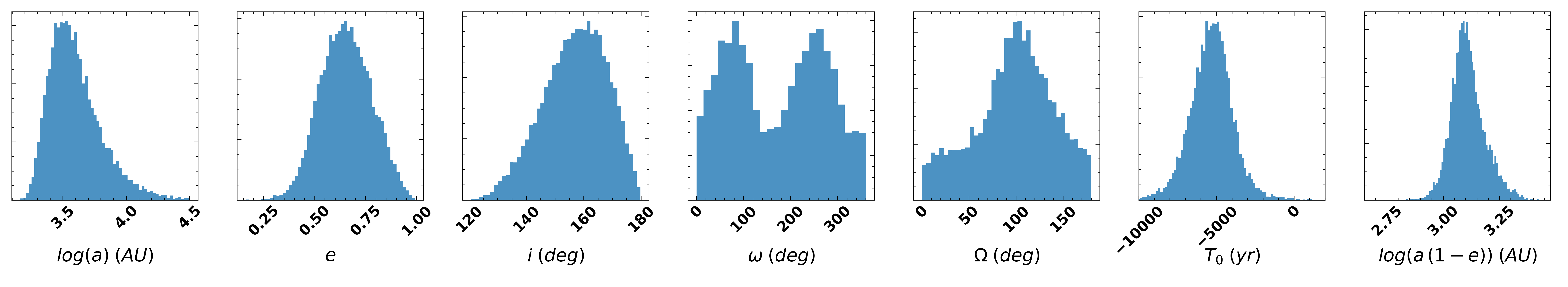}
\caption{\small{ \textbf{Successful fit: GK/GI Tau}. GK Tau hosts a protoplanetary disk with inclination $i = 71 \pm 5 ^\circ$ or $i = 109 \pm 5^\circ$.  The inclination allowed by the \Gaia\ astrometry is limited to 160 $^{+13}_{-10}$ degrees at the 68\% credible interval.  Orbits consistent with GK Tau's disk inclination are ruled out, and the extent of the disk is far smaller than the closest periastron distances, indicating that the wide binary orbit is not influencing the evolution of the disk.
Top:  Selection of 100 orbits from the posterior of the LOFTI fit of GI Tau relative to GK Tau.  Bottom: Posterior distributions of orbital elements for the \Gaia\ position/velocity fit of GI Tau relative to GK Tau.  The long tail of semi-major axis and T$_{\rm{0}}$ has been truncated for clarity.  }}
\label{fig:GKTau_hists}
\end{figure*}

% Let's move this section to Adam's RUWE paper.
%\subsection{HR 858 with \Gaia\}
%\textcolor{blue}{The proximity of a faint companion to a brighter primary does not appear to reduce the effectiveness of this method, provided the quality indicators of Section \ref{sec:intro} indicate both components were robustly fit. HR 858 is a bright ($V$ = 6.4) SpT F4V \citep{Gray2006HR858spt} star hosting 3 transiting planets recently observed by \textit{TESS} \citep{VanderburgHR858}, with a co-moving M-dwarf companion \citep{VanderburgHR858} at 270 au separation. The system is therefore a potential test case for studying for binary-planetary orbital alignment.  HR 858 has a good \Gaia\ DR2 solution (\texttt{parallax\_over\_error} = 446.362, RUWE = 1.063), but the much fainter companion ($\Delta G = 10$) has a poor RUWE (RUWE = 2.283) and significant excess noise (\texttt{astrometric\_excess\_noise} = 0.835, \texttt{astrometric\_excess\_noise\_sig} = 63.402), suggesting either an unresolved binary or the effect of its proximity to a bright star.}

\subsection{GK/GI Tau with \Gaia}

The Taurus pre-main sequence stars GK Tau (SpT = K7; \citealt{Kenyon1995}; M = 0.79 $\pm$ 0.07 M$_{\odot}$ \changed{via dynamical mass}; \citealt{Simon2017}) and GI Tau (SpT = M0.4,  M = $0.53^{+0.09}_{-0.11}$ M$_{\odot}$ \changed{via isochrones}; \citealt{Herczeg&Hillenbrand2014,Akeson2019YoungBinaryDisks}) are a wide separation ($\rho\,=\,$13.2\arcsec, or $\rho = 1700$ AU) binary system in Taurus (\citealt{Hartigan1994,Duchene1999,Kraus&Hillenbrand2009}; distance $d\,=\,129.3 \pm 0.7$ pc).  GI Tau and GK Tau appear to be part of the classically defined Taurus-Auriga star-forming region, with an age of 1--3 Myr \citep{Kraus&Hillenbrand2009}.   
Both GK Tau and GI Tau host gas-rich protoplanetary disks.

\changed{\citet{Long2019TaurusDisks} measure disk inclinations of $i = 40.2^{+5.9}_{-6.2}$ deg and $i = 43.8\pm1.1$ deg, and position angles of PA = $119.9^{+8.9}_{-9.1}$ deg and PA = $143.7^{+1.9}_{-1.6}$ deg for GK Tau and GI Tau respectively\footnote{The inclination of the disk around GK Tau had also been assessed using previous ALMA observations (\citealt{Akeson2014}; \citealt{Simon2017}), but those observations had a much larger beam size (1.20$\times$0.74\arcsec, versus 0.22$\times$0.11\arcsec\, for \citealt{Long2019TaurusDisks}). Given the compact radius of $r_{eff,95\%} = 0.099\arcsec$ reported by \citet{Long2019TaurusDisks}, the previous observations were not strongly constraining. }.  The disk inclinations could also be $139.8 ^{+5.9}_{-6.2}$ deg and $136.2\pm1.1$ deg, as the reported value uses the convention of $i < 90^\circ$. 
}

Table \ref{table:Gaia solutions} shows that \Gaia\ has 
%high S/N parallax measurements and 
low RUWE values for GK Tau (%\texttt{parallax\_over\_error} = 177.959, 
RUWE = 1.089), and GI Tau (%\texttt{parallax\_over\_error} = 166.419, 
RUWE = 1.046).  %There is no astrometric excess noise with GK Tau, consistent with the lack of spectroscopic detection of an unresolved companion by \cite{Kraus&Hillenbrand2009}.  
GI Tau has a small amount of excess noise with low significance\changed{, but the relative velocity is still measured to be non-zero at 10$\sigma$ significance. High resolution imaging has ruled out additional stellar companions to either source down to $\rho \sim 5$ AU \citep{Kraus2011MappingTaurus}. %Gaia DR2 does not report RVs for either component, and while values in the literature are not precise enough to contribute substantively to an orbit, they agree to within $<$2km/s (Hartmann et al. 1986; Nguyen et al. 2012) and hence it is unlikely that either component hosts an additional short-period stellar companion.
}

% GI Tau RV: Nguyen et al. 2012
% GK Tau RV: Hartmann et al. 1986

Figure \ref{fig:GKTau_hists} shows the posterior distributions of the orbital elements resulting from our fit to the \Gaia\ position/velocity alone.  \changed{A binary inclination consistent with either of GK Tau and GI Tau's possible disk inclinations is marginally consistent with the \Gaia\ astrometry,} \changed{though most of the orbital inclination posterior is concentrated in more face-on orientations. In the future, a more precise measurement of the relative RV might allow a more robust test of the mutual inclinations.} \changed{The 95\% lower confidence limit for periastron passage is 890 AU, and for semi-major axis is 1500 AU. Binaries are canonically assumed to externally truncate disks at $\sim$1/2--1/3 of the binary semimajor axis, and very rarely at $r < a/4$ \citep{Artymowicz1994BinaryDiskTruncation}. Both disks are very compact, and \citet{Long2019TaurusDisks} measured $r_{eff,95\%} = 0.190\arcsec$ (25 AU) for GI Tau and $r_{eff,95\%} = 0.099\arcsec$ (13 AU) for GK Tau, so the orbit is not consistent with either disk being truncated at periastron passage.}

\subsection{Kepler-25/ KOI-1803 with \Gaia}

\begin{figure}[!htb]
    \centering
    \includegraphics[width = 0.48\textwidth]{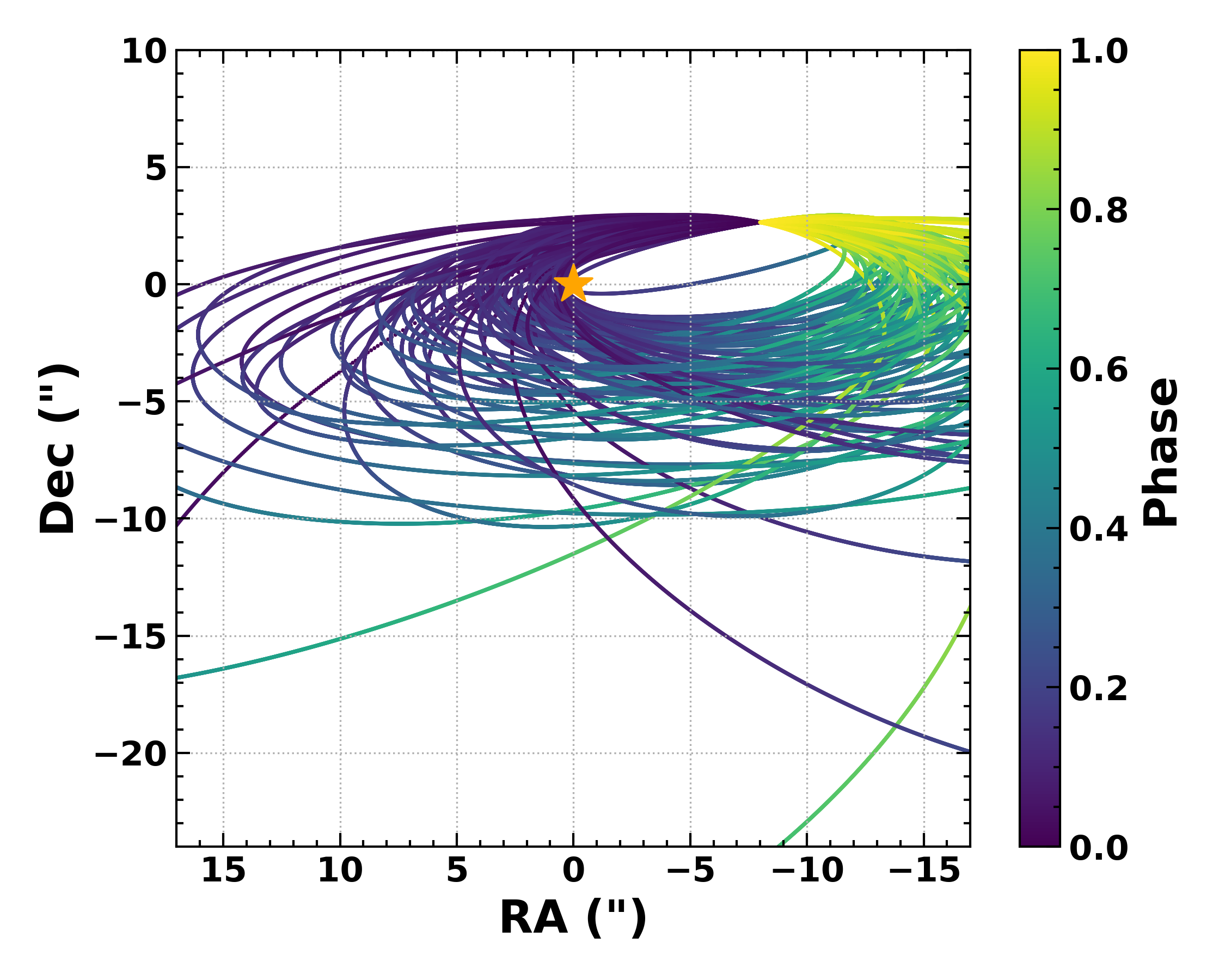}
    \includegraphics[width = 0.48\textwidth]{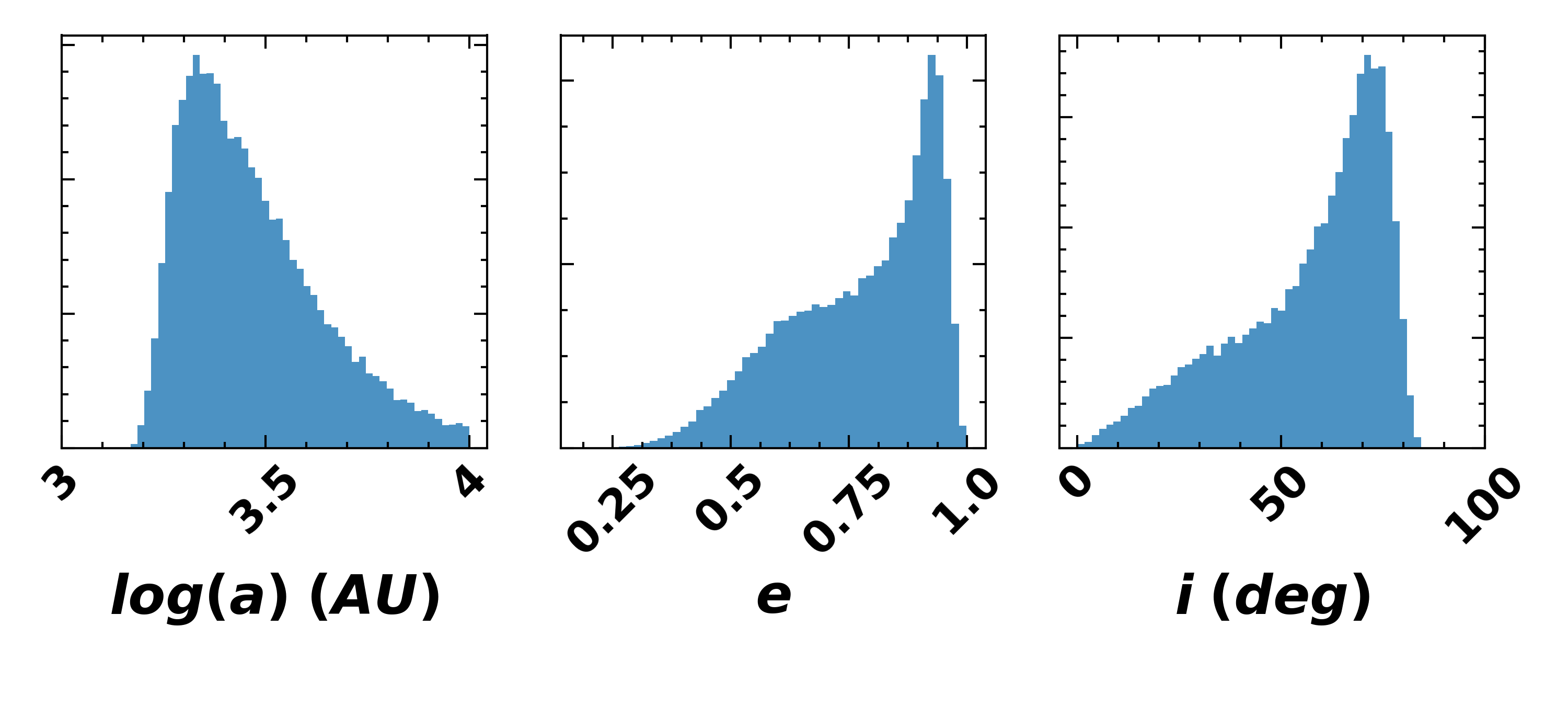}
    \caption{\textbf{Successful fit: Kepler-25/KOI-1803}. Kepler-25 is a transiting planet host, KOI-1803 has a candidate planet signal.  We find that edge-on alignment of the binary with the two transiting planet systems is not allowed by \Gaia\ astrometry. Top: Selection of 100 orbits from the posterior of the LOFTI fit of KOI-1803 relative to Kepler-25. Bottom: Log(semi-major axis), eccentricity, and inclination posterior distributions of KOI-1803/Kepler-25 fit.}
    \label{fig:Kepler-25 orbits}
\end{figure}

The transiting planet host Kepler-25 (KOI-0244;  T$_{\rm eff}~=~6270\pm79$~K,
M~=~1.159$^{+0.040}_{-0.051}$~M$_{\odot}$ \changed{via asterosiesmology}; \citealt{Aguirre2015Asteroseismology}; $\pi = 4.082 \pm 0.023$ mas; \citealt{Gaia2018b}) was found by the Kepler mission to display periodic transit signatures corresponding to two candidate transiting planets. These planets were confirmed via transit-timing variations by \citet{SteffenKepler25TTV2012} as Kepler-25 b ($P = 6.24$ d; R$_{\rm p}$~=~2.6~R$_{\Earth}$) and Kepler-25 c ($P = 12.72$ d; R$_{\rm p}$~=~4.5~R$_{\Earth}$. Radial velocity monitoring by \cite{Marcy2014KeplerPlanetRadii} and \cite{Mills2019KeplerGiantPlanets} subsequently revealed a third, non-transiting giant planet in an outer orbit, Kepler-25 d ($P = 122.4$ d; $M \sin(i) = 0.226 \pm 0.031 M_{Jup}$).

The Kepler Object of Interest KOI-1803 (SpT~=~K1V; \citealt{RoweKeplerValidation2014}; T$_{\rm eff}~=~4979$~K; M~=~0.794$\pm0.026$~M$_{\odot}$ \changed{via isochrones}; \citealt{FultonCalifKeplerSurvey2018}) is located $\rho\,=\,8.42\arcsec$ away from Kepler-25, and is codistant within extremely high precision ($\pi = 4.090 \pm 0.013$ mas; \citealt{Gaia2018b}) and comoving ($\Delta \mu = 0.76 \pm 0.05$ mas yr$^{-1}$). KOI-1803 was also found by the Kepler mission to display three periodic transit signatures (\citealt{RoweKeplerValidation2014}; \citealt{Thompson2018KeplerCand}). Two sets of these apparent transits are ephemeris-matched with the (stronger) signals of Kepler-25, and hence were assessed to be false positives due to PSF overlap in the photometric apertures. However, a third set of transits are deeper than the two false positive signals, yet do not have a counterpart in Kepler-25, and hence indicate a planetary candidate associated with this star (KOI-1803.01; $P = 4.54$ d; R$_{\rm p}$~=~2.3~R$_{\Earth}$) that has not yet been confirmed.

Kepler-25 and KOI-1803 appear to constitute a wide binary system ($\rho \sim 2060$ AU; \changed{this work}) with confirmed planets orbiting the primary star, and one candidate orbiting the secondary star. This wide pair joins the small number of such systems where both components of a binary system have been shown to host transiting planets in edge-on orbits \citep[e.g.][]{Lissauer2014KeplerValidationII}, and hence where the planetary systems might be aligned with each other. The presence of edge-on orbits is not sufficient to confirm coplanarity, as they might be misaligned in $\Omega$. 

However, the Kepler-25/KOI-1803 is unique in being wide enough for both components to possess high-quality Gaia astrometric solutions; if the orbit of the binary components were itself also edge-on, it would provide circumstantial evidence of alignment throughout the system. We therefore performed a fit using the \Gaia\ data for both stars, and show the results in Figure \ref{fig:Kepler-25 orbits}.  

Our fit measures a 95\% confidence upper limit on the inclination of $i < 79.6$\degree, demonstrating that the binary orbit is not seen edge-on and must be misaligned with both planetary systems by $>10$\degree. This misalignment also weakens the case for the two planetary systems being closely aligned with each other, though it does not rule out the possibility. Indeed, a number of young wide binary systems have been shown to host individual circumstellar disks which are not strictly aligned, but are correlated in alignment to within $\la 20$\degree  (e.g., \citealt{Jensen2004ProtoplanetaryDiskAlignment}). This correlated alignment would result in an excess of wide binaries with transiting planets around both stars, above that expected by random pairing, but no quantitative analysis of their occurrence rate has been conducted to date. There are hundreds of KOIs with identified wide binary companions (e.g., \citealt{Deacon2016PanSTARRSKepler}; \citealt{Godoy-Rivera2018KeplerBinaries}) and $\sim$1\% of all Kepler targets host at least one confirmed or candidate planet, so given the absence of more such pairs among the catalog of all KOIs, strict alignment would likely be ruled out by such an analysis and it is possible that the transits of Kepler-25 and KOI-1803 are genuinely coincidental.

% Note, it would be interesting to go back to the Jensen+04 sample and see if Gaia can be used to say much about those systems' binary alignments...

There are hundreds of wide binaries in the Kepler field for which one component hosts a transiting planet while the other does not.  The occurrence rate of planets means it is likely that those stellar companions do host planetary systems that are not aligned.  Given that the odds of observing a transiting planet due to chance random alignment are a few percent, coupled with the misalignment of the binary orbit, this points to chance alignment of Kepler-25 and KOI-1803's planetary orbits rather than the outcome of the formation process.

%%%%%%%%%%%%%%%%%%%%%

\section{Limitations as demonstrated by unsuccessful fits}\label{sec:limitations}
Here we demonstrate where this method might not produce robust orbital posterior distributions, or might be inefficient in exploring those posterior distributions.

\subsection{
Astrometric acceleration during the \Gaia\ time series, such as Gl 896 AB
}
\Gaia\ observations do not offer a superior constraint on an orbit when a significant fraction of the orbit is observed in time series astrometry or when there is acceleration.
GL~896~AB \changed{(a.k.a EQ Peg, BD+19 5116)} is another stellar binary \citep{WirtanenGl896BinaryDiscovery} with long-baseline time series astrometry found in WDS.  The system is a pair of flare stars with M$_A=\,0.3880\pm0.0091 \,$M$_\odot$ and M$_B=\,0.2519\pm0.0076 \,$M$_\odot$, using the Mass-Luminosity relation of \citealt{Mann2019MassLuminosityRelation} and their 2MASS K-band magnitudes \citep{2MASS-Skrutskie2006}.  It is only 6.25 pc distant, and the astrometric observations comprise nearly a quarter of the orbit.

\begin{figure}[!htb]
\centering
\includegraphics[width=0.45\textwidth]{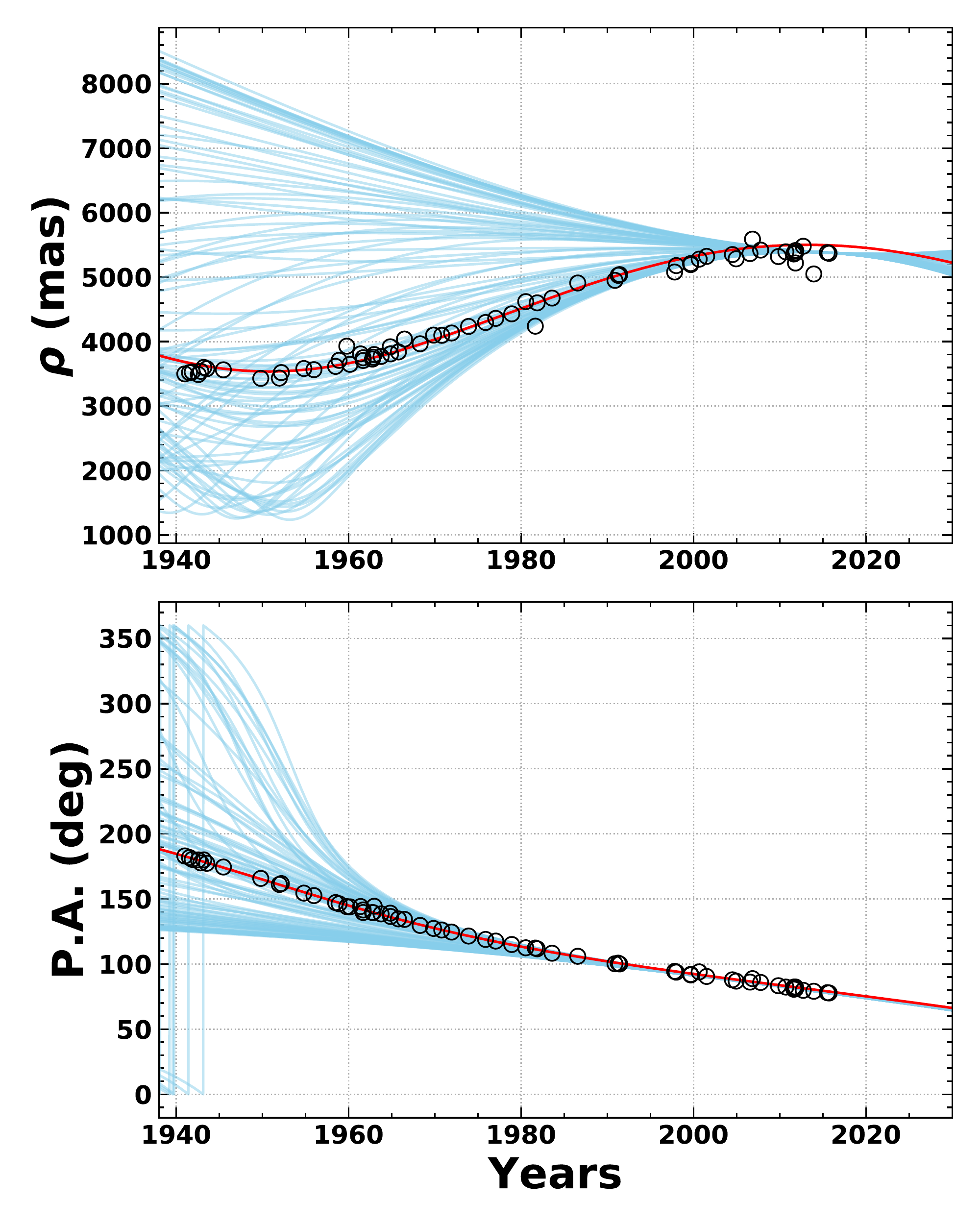}
\caption{\changed{\small{\textbf{Unsuccessful fit: Gl 896}. The \Gaia\-only fit did not constrain the orbit as well as the astrometric fit due to acceleration during the \Gaia\ time series observations. A selection of 100 orbits from the posterior sample from our LOFTI fit to the \Gaia\ measurements for Gl 896B relative to Gl 896A (blue), with the WDS measurements overplotted (open circles, not used in fit) and the orbit of \citealt{Heintz1984Orbitof16VisualBinaries} (red).  The majority of \Gaia\ posterior orbits are not consistent with the long time-series observations.
}}}
\label{fig:GL896 astr fit}
\end{figure}

\changed{\citet{Heintz1984Orbitof16VisualBinaries} determined orbital elements for Gl 896 AB from astrometry spanning four decades.  Table \ref{table:gl896 orbits} displays their orbital elements, and Figure \ref{fig:GL896 astr fit} shows their orbit (red) and all available WDS astrometric observations (open circles) in separation and position angle as a function of time.  \citet{Bower2011GJ896A} measured an acceleration for Gl 896 A relative to B of $(a_\alpha,a_\delta) = (0.3\pm0.1, 3.1\pm0.6)$ mas yr$^{-2}$ using radio interferometry\footnote{The $a_\delta$ value of \citet{Bower2011GJ896A} deviated from that predicted by \citet{Heintz1984Orbitof16VisualBinaries} orbital elements, and they concluded that there is an error in the estimated orbital elements. Nevertheless these orbital elements will suffice for comparisons to our method.}
}

\begin{deluxetable*}{cccccc}[htb!]
\tablecaption{{Comparison of the unsuccessful Gl 896 \Gaia\ LOFTI fit to \citealt{Heintz1984Orbitof16VisualBinaries}}\label{table:gl896 orbits}}
%\tablewidth{0.45\textwidth}
\tablehead{\colhead{Element} & {Heintz 1984}  & \multicolumn{4}{c}{\Gaia\ Position/Velocity fit} \\
\cline{3-6}
\colhead{} & \colhead{} & \colhead{Median} &\colhead{Mode} & \colhead{68.3\% Min CI} & \colhead{95.4\% Min CI} }
\startdata
$a$ (arcsec) & 6.87 & 5.38 & 3.49 & (3.28, 6.78) & (3.27, 16.72)  \\
$P$ (yrs) & 359 & 244 & 132 & (116, 345) & (115, 1337)  \\
$e$ & 0.20 & 0.42 & 0.50 & (0.29, 0.63) & (0.02, 0.64)  \\
$i$ (\degree) & 123.5 & 129 & 127 & (117, 143) & (108, 163) \\
$\omega$ (\degree) & 354.0 & 181  & 264 & (60, 282) & (17, 350) \\
$\Omega$ (\degree) & 82.1 & 96 & 164 & (70, 170) & (12, 180) \\
$T_0$ & 2008 & 1912 & 1945 & (1825, 1966) & (1237, 2015) \\
%\hline
%Periastron (AU) & 5.0$^{+0.9}_{-1.0}$ & 43 & 26 & 25 & (16, 48) & (14, 100) \\
\enddata
%\tablecomments{}
\end{deluxetable*}

Gl 896 AB have high S/N \Gaia\ solutions, but elevated RUWE (A: %\\ \texttt{parallax\_over\_error}~=~1931.8,  
RUWE~=~1.2; B: %\texttt{parallax\_over\_error}~=~1483.2, 
RUWE~=~1.5).  Gl 896 B especially has an RUWE which could indicate 
%the presence of an unresolved companion, or 
that some amount of orbital curvature is observed during the \Gaia\ time-series, which is fit linearly \citep{Lindegren2018}. \changed{\citealt{Bower2011GJ896A} do not find evidence for a short-period companion (M~$>$~1~M$_{\rm{Jup}}$, a~$>$~0.3~AU) around Gl 896 B.} 
Nevertheless, the errors on the \Gaia\ astrometry are small ($\sim$~0.05 mas), enabling the solutions to be used in an orbit fit.  No radial velocity constraint was applied to this system.

\begin{figure}[!htb]
    \centering
    \includegraphics[width=0.45\textwidth]{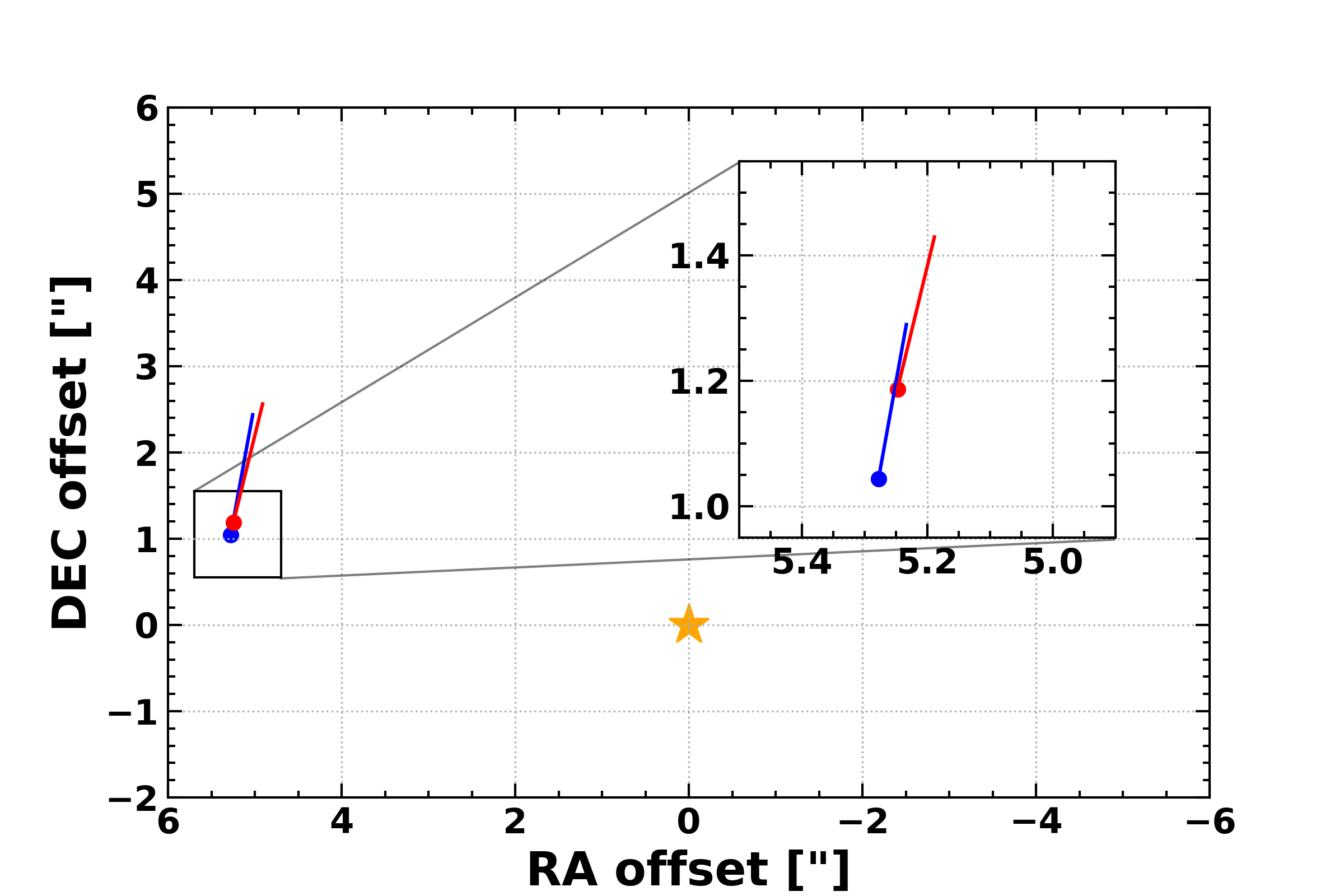}
    \caption{\small{\textbf{Gl 896}. Position and velocity direction of Gl 896B relative to Gl 896A at the beginning (2014 Aug 22 (21:00 UTC), blue) and end (2016 May 23 (11:35 UTC); red) of the \Gaia\ DR2 astrometric time series \citep{Lindegren2018}}.  \changed{Velocity direction was computed from the \citet{Heintz1984Orbitof16VisualBinaries} orbital elements. 
    The velocity direction has changed over the time interval, indicating departure from the assumption of linear motion during the \Gaia\ observations.}  Inset: closer view of position and velocity at beginning and end of \Gaia\ time series, for clarity.}
    \label{fig:GL896 acceleration}
\end{figure}

\changed{Figure \ref{fig:GL896 astr fit} shows a selection of orbits from the posterior of the \Gaia-only fit, as well as the WDS astrometric measurements (black circles) and \citet{Heintz1984Orbitof16VisualBinaries} orbit (red).}  \changed{Table \ref{table:gl896 orbits} shows that the \citet{Heintz1984Orbitof16VisualBinaries} orbit is on the edge of the 68\% minimum credible interval for our posterior distribution.}
Figure \ref{fig:GL896 acceleration} shows the velocity direction at the beginning (2014 Aug 22 (21:00 UTC), blue) and end (2016 May 23 (11:35 UTC); red) of the \Gaia\ DR2 time series \citep{Lindegren2018}, \changed{showing departure from the assumption of linear motion during \Gaia\ DR2 observations}. 
For this system, the \Gaia-only fit did not outperform the time-series astrometric fit, as it had for DS Tuc AB.

%The \Gaia-only fit did return several orbits consistent with the astrometric measurements, but they are only a fraction of the posterior of accepted orbits. 
%Because the time series astrometry covered nearly a quarter of the orbit, acceleration in the plane of the sky can be observed in the astrometry (Figure \ref{fig:GL896 astr fit}). %The posterior of the WDS-only fit has a median acceleration of $\dot\mu~=~2.9^{+0.2}_{-0.4}$~mas\;~yr$^{-2}$ ($ 0.086\, ^{+0.007}_{-0.011}$~km\;~s$^{-1}$~\;yr$^{-1}$) during the \Gaia\ DR2 time series, producing a velocity change that likely was significant to influence the astrometric fit.  

%When enough of the orbit is observed in the \Gaia\ time series to see curvature, the \Gaia\ astrometry alone is no longer suitable for determining the orbit.  
Future \Gaia\ data releases including plane-of-sky acceleration terms will improve the accuracy of orbital element constraint for this and other systems with non-linear motion during the \Gaia\ astrometric observations.

%Additionally we performed an orbit fit combining WDS astrometry and \Gaia\ velocity, shown in Figure \ref{fig:GL896 astr fit} (right).  Naturally, the combined fit is much more constrained than either individually, and also reveals some systematics present in the astrometry (for example, all data points which lie significantly below the orbit results in the 1970's and 1980's came from the same source).  This fit took an exceedingly long time to run with the OFTI rejection sampling algorithm, and would be better served by an MCMC-based algorithm.

\subsection{\changed{Subsystems unresolved by \Gaia}, such as Kepler-444 BC}
The method will not be accurate for systems where one or both objects are unresolved stellar binaries, as in the case of Kepler-444 BC (RUWE$= 16.687$). \changed{Kepler-444 (a.k.a. BD+41 3306, HIP 94931, KOI-3158) consists of a spectroscopic binary, Kepler-444 BC (M$_B$ = 0.29$\pm$0.03 M$_\odot$, M$_C$ = 0.25$\pm$0.03 M$_\odot$ via mass-magnitude relation of \citealt{Delfosse2000MassMagRelation}, as applied by \citealt{Dupuy2016Kepler444}), at 1.8\arcsec separation from the primary Kepler-444 A (\citealt{Carney1983HaloBinaries}; M = 0.76$\pm$0.04 via asteroseismology).  Kepler-444 A hosts five transiting sub-Earth radius planets \citep{Campante2015}. Kepler-444~BC has a high RUWE due to the unresolved binary}, and we find that \Gaia\ astrometry is not reliable for determining the orbit of BC relative to A.  

\begin{deluxetable*}{ccccccc}[htb!]
\tablecaption{{Comparison of the unsuccessful Kepler-444 \Gaia\ LOFTI fit to \citealt{Dupuy2016Kepler444}}\label{table:k444 orbits}}
%\tablewidth{0.45\textwidth}
\tablehead{\colhead{Element} & {Dupuy et.al. 2016}  & \multicolumn{5}{c}{\Gaia\ Position/Velocity fit} \\
\cline{3-7}
\colhead{} & \colhead{} & \colhead{Mean} & \colhead{Std Dev} &\colhead{Mode} & \colhead{68.3\% Min CI} & \colhead{95.4\% Min CI} }
\startdata
$a$ (AU) & 36.7$^{+0.7}_{-0.9}$ & 78.3 & 65.5 & 46.6 & (42.7, 76.0) & (41.1, 171.2)  \\
$P$ (yrs) & 198$^{+8}_{-9}$ & 740 & 1700 & 280 & (250, 600) & (230, 2020) \\
$e$ & 0.864 $\pm$ 0.023 & 0.44 & 0.14 & 0.52 & (0.31, 0.62) & (0.18, 0.67) \\
$i$ (\degree) & 90.4$^{+3.4}_{-3.6}$ & 133.4 & 11.1 & 138.9 & (124.8, 148.5) & (112.3, 151.7)\\
$\omega$ (\degree) & 342.8$^{+3.2}_{-2.6}$ & 6  & 89 & -80 & (-90, 107) & (-150, 174)\\
$\Omega$ (\degree) & 73.1 $\pm$ 0.9 & 77 & 45 & 121 & (-57, 132) & (-13, 134)\\
$T_0$ (JD) & 2488500 $\pm$ 900 & 2328900 & 376400 & 2393700 & (2367900, 2402600) & (2126200, 2408900)\\
%\hline
Periastron (AU) & 5.0$^{+0.9}_{-1.0}$ & 43 & 26 & 25 & (16, 48) & (14, 100) \\
\enddata
%\tablecomments{}
\end{deluxetable*}

\begin{figure*}
\centering
\includegraphics[width=0.95\textwidth]{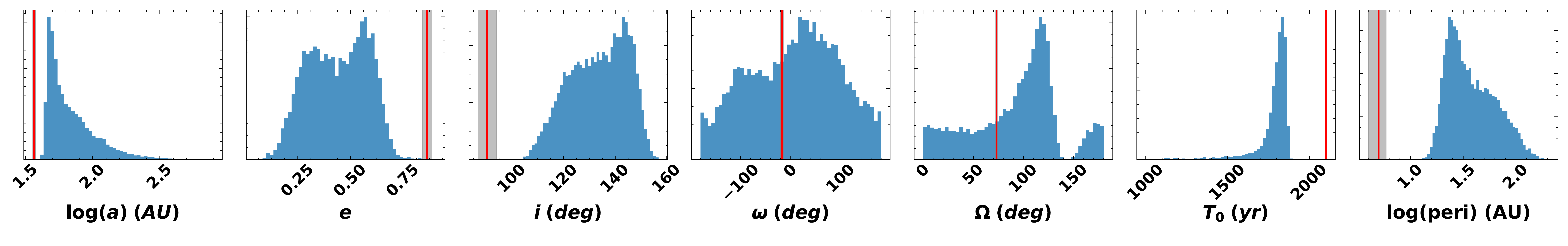}
\caption{\small{\textbf{Unsuccessful fit: Kepler-444}. The \Gaia\ astrometry is not sufficient to accurately constrain the orbit of this system due to one component being a close binary unresolved by \Gaia\.  Posterior distributions of orbital elements for the LOFTI fit of Kepler-444 are shown. The mean values reported in D16 are marked by red lines, error in grey. The LOFTI posteriors are not consistent with the established orbit for the system. }}
\label{fig:GKTau}
\end{figure*}

We performed an orbit fit anyway to demonstrate the effect of the \changed{unresolved binary} on the orbit determination. Table \ref{table:k444 orbits} displays the results of our \Gaia\ linear velocity fit as compared to the established orbit of \cite{Dupuy2016Kepler444}.  For the majority of parameters the established value does not even fall within the 95\% confidence interval for our posterior. 

\changed{The accuracy of the \Gaia\ orbit fitting technique will be impacted when wide binaries contain subsystems.  The extent to which subsystems may bias \Gaia\ measurements remains unknown, but is partially captured in the RUWE parameter.  We therefore recommend careful consideration of the RUWE of components when applying this technique}.

\subsection{\changed{Possible source confusion due to small separations}, such as RW Aurigae}

RW Aurigae is a T-Tauri system with at least two components, RW Aurigae A and B, with a separation of $\rho = 1.5$\arcsec \citep{Duchene1999}.  \cite{Rodriguez2018} imaged disks around both A and B with ALMA, with disk inclinations $i_A = 57.68\degree \pm 0.86\degree$ and $i_B = 72.08\degree \pm 7.98\degree$ in ALMA Band 7, and found evidence that the disks had been disrupted by a possible stellar flyby.  
The two components are resolved in \Gaia\ DR2, making this an interesting case for studying the alignment between the stellar orbit and the disks.

The 1.5\arcsec\ separation between the components is much closer than other systems we explored here.  Table~\ref{table:Gaia solutions} shows that RW Aurigae B has a high RUWE value (A:% \texttt{parallax\_over\_error} = 92.002, 
RUWE = 1.774; B: %\texttt{parallax\_over\_error} = 7.296,
RUWE = 28.904).  \changed{The high RUWE for component B could be due to an unresolved companion, as with Kepler-444. However, \citealt{Kraus2011MappingTaurus} ruled out companions with contrasts down to $\Delta K' \sim$ 3.5 mags at 40 mas, and $\Delta K' \sim$ 1.5 mags down to 20 mas in non-redundant aperture masked (NRM) imaging.  To induce sufficient noise that B would have RUWE $\sim$ 30, a companion around RW Aurigae B would likely be wide and bright enough to be detectable by NRM (Kraus et al. in prep).
}

\changed{We suggest that the binary separation might introduce confusion in the astrometric solution.  %In addition to the potential for the wings of the PSF to overlap out to several $\lambda$/D, 
The orientation of the two components relative to the scan direction could cause projection effects when projected to the line spread function (LSF) \citep{GaiaCollaboration2016TheMission}.  If the components are oriented nearly perpendicular to the scan direction, their profiles might blend and overlap when projected to the 1-D LSF, and confuse the astrometric solution.  Kraus et al. (in prep) have found that for binary systems of roughly equal brightness, those with projected separations below $\rho \sim$1.5\arcsec\ show elevated RUWE values for both components, especially the secondary.  This effect could be the source of the elevated RUWE values for RW Aurigae.  % Cool. See the plot I just sent - that might provide context for whether this can be the cause for B's higher RUWE, so maybe worth adding another sentence on that? -ALK
}
%It is possible that the overlap of the point spread function of the two components could cause the apparent photocenter to shift over time in a way that is discrepant with their orbital motion.

We attempted an orbit fit to the \Gaia\ data.  The large errors on astrometric parameters meant that orbital parameters were poorly constrained by the data, \changed{and did not produce reliable results.  %The stellar orbital inclination was constrained to $ i = 123 ^{+20}_{-24} $ degrees, \changed{which could be consistent with RW Aurigae A's disk as the disk inclination was determined in the [0,90]$^\circ$ range}.  However as discussed above, this result should not be relied upon. 
Improved astrometry is needed to apply LOFTI.}

\section{Discussion and Conclusion}

We have shown that \Gaia\ astrometry alone \changed{can be used} to provide \changed{scientifically interesting} constraints on orbital elements in several test cases.  The suitability of \Gaia\ astrometry for any particular binary system must be carefully assessed before being used as the sole observational constraint for stellar orbit fitting. Binaries for which both components have RUWE $\approx$1.0 are good candidates for this technique.  Care should be taken in cases where the orbital period is short enough to influence the \Gaia\ astrometric result, where there might be unresolved \changed{inner} companions, or where a resolved yet close separation binary may influence the \Gaia\ astrometry.  \changed{Careful consideration of the RUWE value of both components must be undertaken before relying on orbit fitting results from \Gaia\ measurements alone.}

The lack of the requirement for long-term astrometric monitoring to constrain orbital elements enables new investigations on many topics in binary star science.  Co-planarity between binary and planetary orbits, or binary orbit and protoplanetary disk, is easily determined, as in the case of DS Tuc, or ruled out, as with Kepler-25 and KOI-1803.  The potential for the binary orbit to have influenced development of a disk or planetary system can be ruled out in systems with low eccentricity or wide periastron distances, as with GK/GI Tau.

\changed{While \Gaia\ astrometry alone does not fully determine an orbit, we have already demonstrated how the ease of access and accuracy of \Gaia\ measurements can readily contribute to study of binary system dynamics and star and planet formation.
}
%The true power of this technique will come with future \Gaia\ data releases. Increasing astrometric accuracy for an increasing number of systems of wider diversity, new astrometric terms including plane-of-the-sky accelerations, and increasing number of radial velocity measurements will allow wider application of \Gaia for placing scientifically valuable limits on orbital parameters

\acknowledgments

The authors wish to thank Eric Gaidos, Saul Rappaport, Kaitlin Kratter, and Renu Malhotra for helpful conversations and suggestions.  We thank the anonymous referee for helpful feedback and suggestions.

L.A.P. was supported by a NASA/Keck Data Analysis Grant and by the McDonald Observatory Board of Visitors, the Cox Endowment Fund of the UT-Austin Department of Astronomy, the Barry Goldwater Scholarship and Excellence in Education Foundation, the Astronaut Scholarship Foundation, and the NSF Graduate Research Fellowship.  This material is based upon work supported by the National Science Foundation Graduate
Research Fellowship Program under Grant No. DGE-1746060. Any opinions,
findings, and conclusions or recommendations expressed in this material are those of the author(s) and do not necessarily reflect the views of the National Science Foundation.

T.J.D.~acknowledges research support from Gemini Observatory, which is operated by the Association of Universities for Research in Astronomy, Inc., on behalf of the international Gemini partnership of Argentina, Brazil, Canada, Chile, the Republic of Korea, and the United States of America.

This work has made use of data from the European Space Agency (ESA) mission
{\it Gaia} (\url{https://www.cosmos.esa.int/gaia}), processed by the {\it Gaia}
Data Processing and Analysis Consortium (DPAC,
\url{https://www.cosmos.esa.int/web/gaia/dpac/consortium}). Funding for the DPAC
has been provided by national institutions, in particular the institutions
participating in the {\it Gaia} Multilateral Agreement.

This research has made use of the Washington Double Star Catalog maintained at the U.S. Naval Observatory.

This publication makes use of data products from the Two Micron All Sky Survey, which is a joint project of the University of Massachusetts and the Infrared Processing and Analysis Center/California Institute of Technology, funded by the National Aeronautics and Space Administration and the National Science Foundation.

\software{Numpy \citep{numpy}, Astropy \citep{astropy:2018}, Matplotlib \citep{Hunter:2007Matplotlib}}

%%%%%%%%%%%%%%%%%%%%%%%%%%%%% Appendix %%%%%%%%
\appendix

\section{Equations for position, velocity, and acceleration given orbital elements}

Here we present the equations we have used to compute predicted position, velocity, and acceleration for a trial orbit in our LOFTI orbit fits.  

For fitting orbits to stellar binaries in \textit{Gaia}, we reduce the two-body system to the relative motion of one mass-less point particle around a central object of mass equal to the total system mass.  Taking the central body of a 2-body Keplerian orbit to be at the origin of the 3-d cartesian coordinate system, the position of the orbiting body is given by the coordinates (X,Y,Z), where +X is the reference direction, equal to +Declination in the on-sky coordinates.  +Y is the +RA direction, and +Z is the line of sight direction towards the observer. This is the coordinate system presented in \cite{Murray2010}, from which we base our derivation. We note that this is different from the typical radial velocity convention that is often used elsewhere, in which -Z is toward the observer.  % (Note - this is a different convention than the typical radial velocity convention, for which -Z is towards to observer)

The orbital elements are: $a$ (semi-major axis) [and thus \changed{$P$} (period - derived from Kepler's 3$^{rd}$ law \changed{utilizing the total system mass and parallactic distance})], $t_0$ (time of periastron passage), $e$ (eccentricity), $i$ (inclination), $\omega$ (argument of periastron - angle from ascending node to periapse), and $\Omega$ (longitude of periastron - angle of location of ascending node from reference direction). %Other symbols are often used for these variables, such as $P$ rather than $T$ to denote period, but we use the symbols as defined by \citet{Murray2010} for consistency.  %(Note - other symbols are often used from these variables [for example, P rather than T for period].  We use the symbols as they are defined in \cite{Murray2010} for consistency).

\subsection{Positions}

\cite{Murray2010} Equations 53, 54, and 55 derive the following formulae for projecting orbital elements onto the plane of the sky:
\begin{equation}
    X = r[\cos\Omega\cos(\omega + f) - \sin\Omega\sin(\omega + f)\cos i]
\end{equation}
\begin{equation}
    Y = r[\sin\Omega\cos(\omega + f) - \cos\Omega\sin(\omega + f)\cos i]
\end{equation}
\begin{equation}
    Z = r \sin(\omega + f)\sin i
\end{equation}

+X and +Y correspond to the observed +$\Delta$Dec ($\Delta\delta$) and +$\Delta$RA ($\Delta\alpha$) respectively between the orbiting body and central object. In this system +Z is defined toward the observer, contrary to the more commonly used radial velocity convention.

The radius of the orbiting body in the orbital plane is denoted as $r$, and is given by:
\begin{equation}
    r = \frac{a(1-e^2)}{1+e \cos f}
\end{equation}

The true anomaly is denoted as $f$, and is given by solving Kepler's equation at the observation date (for \textit{Gaia} DR2 this is 2015.5):
\begin{equation}
    M = \frac{2 \pi}{P} \big(t - t_o \big)
\end{equation}
\begin{equation}
    \changed{M = E - e \sin E}
\end{equation}
which is a transcendental equation which must be solved numerically.  In this case, $P$ is derived from Kepler's 3rd law as $T = \sqrt{\frac{4 \pi^2 a^3}{\mu}}$, where $\mu = G (m_1 + m_2)$.  The true anomaly then is given by:
\begin{equation}
    f =2\,\arctan \left(\,\sqrt {\frac{1+e\,}{1-e\,}} \tan {\frac{E}{2}}\,\right)
\end{equation}

\subsection{Velocities}
\cite{Murray2010} derive in Equation 63 the formula for velocity in the Z direction (radial velocity) as:
\begin{equation}
    \dot Z = \dot r \sin(\omega + f) \sin i + r \dot f \cos(\omega + f) \sin i
\end{equation}
where $\dot{Z}$ is the time derivative of Z.  In the equations above, only $r$, $\dot{r}$, and $f$ vary with time.

The time derivatives of X and Y give the velocity in the X and Y direction, which corresponds to proper motion in the Dec and RA directions respectively ($\mu_\delta$ and $\mu_\alpha$).

\begin{equation}
\begin{multlined}
    \dot X = \dot r \, \big[\, \cos\Omega \cos(\omega + f) - \sin\Omega\sin(\omega + f)\cos i\,\big] + \\ r \dot f \,\big[\, -\cos\Omega\sin(\omega + f) - \sin\Omega\cos(\omega + f)\cos i \,\big]
\end{multlined}
\end{equation}
\begin{equation}
\begin{multlined}
    \dot Y = \dot r \,\big[ \sin\Omega \cos(\omega + f) + \cos\Omega\sin(\omega + f)\cos i\,\big] + \\ r \dot f \,\big[-\sin\Omega\sin(\omega + f) + \cos\Omega\cos(\omega + f)\cos i\,\big]
\end{multlined}
\end{equation}
where $\dot r$ and $r \dot f$ are the time rate of change of separation and angular distance from the focus of the ellipse (the central body).  
Equations 31 and 32 in \cite{Murray2010} define $\dot r$ and $r \dot f$ in terms of $a$ , $e$ , and $f$:
\begin{equation}
    \dot r = \frac{n a}{\sqrt{1-e^2}} e \sin f
\end{equation}
\begin{equation}
    r \dot f = \frac{n a}{\sqrt{1-e^2}} \big(1+e\cos f \big) 
\end{equation}
where $n = \frac{2\pi}{T}$.

And the final position and velocity equations become:
\begin{equation}
\begin{multlined}
    X =  \frac{a(1-e^2)}{1+e \cos f} (\cos\Omega\cos(\omega + f) - \sin\Omega\sin(\omega + f)\cos i)  = \Delta \delta
\end{multlined}
\end{equation}

\begin{equation}
\begin{multlined}
    Y =  \frac{a(1-e^2)}{1+e \cos f} (\sin\Omega\cos(\omega + f) - \cos\Omega\sin(\omega + f)\cos i)  = \Delta \alpha
\end{multlined}
\end{equation}

\begin{equation}
    Z =  \frac{a(1-e^2)}{1+e \cos f} \sin(\omega + f)\sin i 
\end{equation}

\begin{equation}
\begin{multlined}
    \dot X = \frac{n a}{\sqrt{1-e^2}} \;[\; e \sin f (\cos\Omega \cos(\omega + f) - \sin\Omega\sin(\omega + f)\cos i) \;  + \\ \big(1+e\cos f \big) (-\cos\Omega\sin(\omega + f) - \sin\Omega\cos(\omega + f)\cos i) \;] = \mu_\delta
\end{multlined}
\end{equation}

\begin{equation}
\begin{multlined}
    \dot Y = \frac{n a}{\sqrt{1-e^2}} \;[\; e \sin f (\sin\Omega \cos(\omega + f) + \cos\Omega\sin(\omega + f)\cos i) \; + \\ \big(1+e\cos f \big) (-\sin\Omega\sin(\omega + f) + \cos\Omega\cos(\omega + f)\cos i) \;] = \mu_\alpha
\end{multlined}
\end{equation}

\begin{equation}
\begin{multlined}
    \dot Z = \frac{n a}{\sqrt{1-e^2}} \big[ e \sin f \sin(\omega + f) \sin i +\big(1+e\cos f \big) \cos(\omega + f) \sin i \big] = v_{radial} %RV
\end{multlined}
\end{equation}

\subsection{Accelerations}
\changed{Future \Gaia\ data releases will include terms for accelerations in the plane of sky.  For completeness, we derive here equations for $\ddot{X}$, $\ddot{Y}$, $\ddot{Z}$ in terms of orbital elements, anticipating the use of these measurements in future orbit fitting with \Gaia.}

Beginning with Equations (A8)-(A10), we derive the second time derivative for X, Y, and Z position as

\begin{equation}
\begin{multlined}
    \ddot{X} =  (\ddot{r} - r \dot{f}^2 )\,\big[\, \cos \Omega \cos(\omega + f) - \sin \Omega \sin(\omega + f) \,  \cos i\, \big] \;+ \\ (-2 \dot{r} \dot{f} - r \ddot{f})\,\big[\, \cos \Omega \sin (\omega+f) + \sin \Omega \cos(\omega+f) \cos i \, \big]
\end{multlined}
\end{equation}

\begin{equation}
\begin{multlined}
    \ddot{Y} = ( \ddot{r} - r \dot{f}^2 )\, \big[\, \sin \Omega \cos(\omega + f) + \cos \Omega \sin(\omega + f) \cos i \, \big] \;+ \\ (2 \dot{r} \dot{f} + r \ddot{f})\, \big[\, \sin \Omega \sin (\omega+f) + \cos \Omega \cos(\omega+f) \cos i\, \big]
\end{multlined}
\end{equation}

\begin{equation}
\begin{multlined}
    \ddot{Z} = \sin{i}\, \big[ \,(\ddot{r} - r \dot{f}^2\,)\,\sin{(\omega + f)} + (2 \dot{r} \dot{f} + r \ddot{f} ) \, \cos(\omega + f) \, \big]
\end{multlined}
\end{equation} 

\cite{Klioner2016} gives two expressions for $\dot{E}$:
\begin{equation}
    \dot{E} = \frac{n}{1-e\cos E}
\end{equation}
\begin{equation}
    \dot{E} = \frac{a n}{r} = \frac{n\; (1+e \cos{f})}{1-e^2}
\end{equation}

Thus we derive from Equation (A22):
\begin{equation}
    \ddot{E} = \frac{-\;n\;e\;\sin{E}}{(1-e \cos{E})^2} \; \dot{E} = \frac{n^2\;e}{(1-e\cos{E})^2}\; \frac{\sin f}{\sqrt{1-e^2}}
\end{equation}

Or from (A23):
\begin{equation}
    \ddot{E} = \frac{-n\;e\;\sin{f}}{1-e^2}\; \dot{f}
\end{equation}

From Equation (A12), we find that
\begin{equation}
    \dot{f} = \frac{n \sqrt{1-e^2}}{(1-e\cos E)^2} = \dot{E} \frac{\sqrt{1-e^2}}{1-e \cos E} = \dot{E} \; \frac{\sin f}{\sin E}
\end{equation}

Where $\sin f = \frac{\sqrt{1-e^2} \sin E}{1-e \cos E}$.  %Which is kind of a nice result.

\changed{Rewriting (A11) as:}
\begin{equation}
    \dot{r} = a\;e\;\dot{E}\;\sin{E}\;
\end{equation}

so
\begin{equation}
    \ddot{r} = a\,e\,\cos{E}\,\dot{E}^2\; + \;a\,e\,\sin{E}\;\ddot{E}
\end{equation}

And from Equation (A26) we derive:
\begin{equation}
    \ddot{f} = \ddot{E}\;\frac{\sqrt{1-e^2}}{1-e\cos E} + \dot{E}^2\, \frac{e \sqrt{1-e^2}\sin E}{(1-e\cos E)^2}
\end{equation}

Which reduces to:
\begin{equation}
    \ddot{f} = \ddot{E}\;\frac{\sin f}{\sin E} + \dot{E}^2\, \frac{e \sin{f}}{1-e\cos E}
\end{equation}

This allows the calculation of all needed variables for computing $\ddot{X}$, $\ddot{Y}$, and $\ddot{Z}$.

\section{The LOFTI python tool}

Documentation, tutorials, and illustration of functions are provided at the
\href{https://github.com/logan-pearce/lofti\_gaiaDR2}{\texttt{lofti\_gaiaDR2} GitHub repository}.  Readers are directed to that repository for a more in-depth discussion of functionality.

Briefly, the \texttt{lofti\_gaiaDR2} python tool wraps the functionality of the LOFTI method described in Section 2.1 into a minimal python user interface.  The user inputs the \Gaia\ DR2 source id numbers, which can be found at the \href{https://gea.esac.esa.int/archive/}{\Gaia\ archive}, the mass and uncertainties for each component, and a minimum number of desired orbits, into the \texttt{fitorbit} module.  \texttt{fitorbit} queries the Gaia repository for the observational constraints, and runs trial orbits for the second source relative to the first source until the minimum desired number of orbits are accepted.  

\begin{figure}[!htb]
    \centering
    \includegraphics[width = 0.48\textwidth]{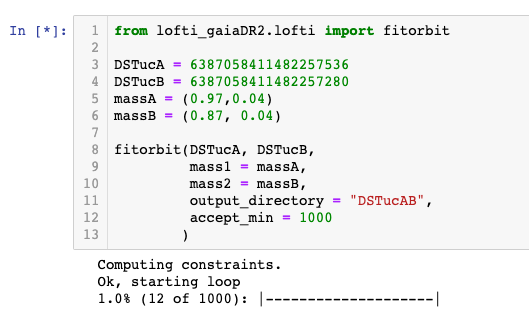}
    \caption{Example usage of \texttt{lofti.fitorbit} to use \Gaia\ archive measurements to fit the orbit of DS Tuc B relative to DS Tuc A.  The user inputs the two \Gaia\ source ids, masses and uncertainties, an output directory to store the accepted orbits, and an minimum number of accepted orbits desired. The minimum orbits desired was set to 1000 for this demonstration, however to obtain a statistically meaningful result that samples the multi-dimensional posterior well, a minimum of $\ge$10$^{5}$ orbits is recommended.  The function queries the \Gaia\ archive to obtain the measurement constraints, and runs trial orbits, updating the user via the progress bar, until the minimum number of orbits is accepted.}
    \label{fig:lofti_example}
\end{figure}

Figure \ref{fig:lofti_example} shows example input and output for \texttt{fitorbit} using the \Gaia\ source ids for DS Tuc A and B given in Table \ref{table:Gaia solutions}, and masses from \citealt{Newton2019DSTuc}.  The source ids, masses, directory for storing results, and minimum desired orbits are input to \texttt{fitorbit}.  If masses are omitted, the function will prompt the user to supply them.
\texttt{fitorbit} runs trial orbits using the priors and acceptance criteria given in Section 2.1, writes accepted orbital parameters to a file contained in the specified output directory, and reports progress towards desired orbits through the progress bar.  The function will provide a warning if either object has RUWE $\ge$ 1.2.  \changed{If the \Gaia\ archive includes radial velocity measurements for both objects, it will be automatically incorporated into the fit.  At this time, no other user inputs, such as time-series astrometry or independently determined radial velocities, are accepted. The function fits using \Gaia\ measurements alone.}

\texttt{lofti\_gaiaDR2} includes the plotting function \texttt{makeplots}.  The \texttt{makeplots} function produces plots of the \texttt{fitorbit} output like the ones included in this paper.  Specifically, calling \texttt{makeplots} for the output file from \texttt{fitorbit} will produce sky plots like Figure \ref{fig:dstuc sky orbits}, distributions of computed positions ($X,Y,Z$), velocities ($\dot X,\dot Y,\dot Z$), and accelerations ($\ddot X,\ddot Y,\ddot Z$) of the posterior orbits, a 3-dimensional orbit plot of selected orbits, 1-dimensional posterior distributions of orbital parameters like Figure \ref{fig:dstuc fit hists}, and a statistics file describing the distributions of orbital parameters.  Each of these outputs can be toggled on and off through keywords in the function call.  \changed{Users can choose to limit $\Omega$ or $\omega$ to [0,180] deg interval if no RV measurements were used, plot semi-major axis or periastron in log scale, or truncate the long tail of the semi-major axis distribution through function keywords.}
The outputs are written to the user specified directory.

Python packages numpy, matplotlib, astropy, astroquery, and pickle are dependencies for the \texttt{lofti\_gaiaDR2} python tool.  \texttt{lofti\_gaiaDR2} can be installed via pip by calling \texttt{pip install lofti\_gaiaDR2} or through the \href{https://github.com/logan-pearce/lofti\_gaia}{the \texttt{lofti\_gaia}} github repository.


\bibliography{ref,references}

\end{document}